\documentclass{aa}
\usepackage[utf8]{inputenc} 
\usepackage[colorlinks=true, citecolor=blue, linkcolor=blue]{hyperref}
\usepackage{cases}
\usepackage[english]{babel}
\usepackage{multicol}
\usepackage{amsmath}
\usepackage{amsfonts}
\usepackage{dsfont}
\usepackage{amsxtra}
\usepackage{amssymb}
\usepackage{upgreek} 
\usepackage{changes}
\usepackage{multirow}
\usepackage{url}
\usepackage{float}

\usepackage{graphicx,epsfig}
\usepackage{dcolumn}

\usepackage{xspace} 

\usepackage{color}
\usepackage{xcolor}

\usepackage[switch]{lineno}
\usepackage{lipsum}

\begin{document}

   \title{Evidence for hybrid gamma-ray emission from the supernova remnant G150.3+4.5}


   \author{Yuan Li\thanks{yuanlss17@sjtu.edu.cn}
          \inst{1,2}
          \and
          Siming Liu\thanks{liusm@swjtu.edu.cn}
          \inst{3}
          \and
          Gwenael Giacinti\inst{1,2}\thanks{gwenael.giacinti@sjtu.edu.cn}
          }

   \institute{Tsung-Dao Lee Institute, Shanghai Jiao Tong University, Shanghai 201210, PRC
        \and
              {School of Physics and Astronomy, Shanghai Jiao Tong University, Shanghai 200240, PRC}
         \and
             School of Physical Science and Technology, Southwest Jiaotong University, Chengdu 610031, PRC\\
             }

  \abstract 
   {Supernova remnant (SNR) G150.3+4.5 was first identified in radio and has a hard GeV spectrum with a $\sim 1.5^\circ$ radius. Radio observations revealed a bright arc with an index of $\sim -0.40$ in contrast to the index of $\sim -0.69$ for the other part. This arc is coincident with the point-like \emph{Fermi} source 4FGL J0426.5+5434 and KM2A source 1LHAASO J0428+5531. The rest of the SNR however has a hard GeV spectrum and a soft TeV spectrum, implying a spectral cutoff or break near 1 TeV. Since there is no X-ray counterpart and no pulse signal detected, the $\gamma$-ray emission mechanism from the SNR and the point-like source are puzzling. We reanalyse the $\gamma$-ray emission using 14 yr data recorded by \emph{Fermi} Large Area Telescope and find that the spectrum of the northern half-sphere is compatible with a broken power-law with a break at 146 $\pm$ 11 GeV and photon indices of $\Gamma_{\rm{NorthLobe}}$ =$1.54\pm0.04_{\rm{stat}}\pm0.07_{\rm{syst}}$ ($2.28\pm0.08_{\rm{stat}}\pm0.12_{\rm{syst}}$) below (above) the break, the southern half-sphere can be described well with a single power-law with $\Gamma_{\rm{SouthLobe}}$ =$1.95\pm0.07_{\rm{stat}}\pm0.09_{\rm{syst}}$. Since the southern half-sphere is well correlated with CO emission, we propose that the $\gamma$-ray emission of the northern half-sphere is dominated by relativistic electrons via the inverse-Compton processes, while the southern half-sphere is dominated by cosmic rays via the hadronic processes. 4FGL J0426.5+5434 can result from illumination of a cloud by escaping cosmic rays or recent shock-cloud interaction. Observations from LHAASO-KM2A then favor the possibility of a cosmic-ray PeVatron candidate, while leptonic scenarios can not be ruled out. Further multi-wavelength observations are warranted to confirm the hadronic nature of 1LHAASO J4028+5531.}


   \maketitle

\section{Introduction}

Supernova remnants (SNRs) are believed as the most probable locations for accelerating Galactic cosmic rays up to the energy of the spectral knee \citep{Ginzburg&Syrovatskii1964, Hillas2005}. According to the diffusive shock acceleration (DSA) theory, self-produced magnetic turbulence may scatter CRs as they are accelerated at the shock of SNRs. 
Additionally, the highest-energy CRs near shock precursor have a poor ability to effectively create their own turbulence and tend to escape from the SNR. These escaped CRs might interact with surrounding dense materials like molecular clouds (MCs) through pp collisions leading to significant brightness in the $\gamma$-ray range. The decay of neutral pions produced by inelastic collisions between accelerated protons and thick gases in MCs is frequently used to explain $\gamma$-ray emissions in the GeV range from bright SNRs. Notably, peculiar characteristics connected to $\pi^0$ decay can be seen in several $\gamma$-ray spectra with the \emph{Fermi} Large Area Telescope (\emph{Fermi}-LAT), such as W44 \citep{uchiyama2012fermi, peron2020gamma}, W28 \citep{aharonian2008discovery, li2010gamma, hanabata2014detailed}, which is thought to be the clearest indication of relativistic nuclei acceleration in SNRs. The flux of $\gamma$-rays depends on the quantity of nuclear cosmic rays released and the diffusion coefficient within the interstellar medium \citep[ISM;][]{aharonian1996emissivity,gabici2009,aharonian2004,marrero2008}, and analyses of the spectra of $\gamma$-rays from different regions of SNRs offer valuable insights into the diffusion process and even constrain the energy dependence of the diffusion coefficient in the interstellar medium \citep{aharonian1996emissivity,ohira2011}.

PeVatrons refer to astrophysical sources that can accelerate cosmic rays (CRs) to the PeV energy range around the knee of the CR energy distribution. Many ultra-high-energy (UHE) $\gamma$-ray sources have been recently detected owing to unprecedented sensitivities of $\gamma$-ray observatories like the Large High Altitude Air Shower Observatory (LHAASO) \citep{2021Natur.594...33C} and the High Altitude Water Cherenkov (HAWC) observatory \citep{2023NIMPA105268253A,2021NatAs...5..465A}. Most of these sources are associated with pulsars, and their UHE gamma rays are likely produced by UHE electrons/positrons via the inverse Compton scattering of low energy photons in the background \citep{2023arXiv230517030C}. Finding hadronic UHE $\gamma$-ray sources (PeVatrons) plays an essential role in addressing the origin of the most energetic Galactic CRs. Although the origin of PeV CRs remains obscure, several sources, including the 
Boomerang supernova remnant (SNR) \citep{2021NatAs...5..460T}, the Cygnus region \citep{2021NatAs...5..465A, 2024SciBu..69..449L},
and LHAASO J2108+5157 \citep{2021ApJ...919L..22C}, have been considered as important sources of PeV CRs. 
In particular, LHAASO J2108+5157 has been shown to be associated with molecular clouds, 
indicating a hadronic scenario for its $\gamma$-ray emission \citep{2023PASJ...75..546D,2023A&A...675L...5D}. Nevertheless, the exact nature of their UHE $\gamma$-ray emission is still a matter of debate.

G150.3+4.5 was first observed as an SNR candidate with faint radio emission during the Canadian Galactic Plane Survey (CGPS) given by \citet{2014A&A...566A..76G}. Subsequent detailed observation in radio band reported in \citet{2014A&A...567A..59G} shows that the radio spectral indices of the southeastern and western arcs differ significantly with $\alpha_e$ = $-0.40 \pm 0.17$ and $\alpha_w$ = $-0.69 \pm 0.24$, respectively ($S_\nu \propto \nu^{\alpha_{\rm i}}$). 
Its overall GeV spectrum is however hard.
Recent $\gamma$-ray analysis by 
\citet{2020A&A...643A..28D} suggested a leptonic scenario for the $\gamma$-ray emission and the soft point source 4FGL J0426.5+5434 might be associated with a powerful pulsar, even though no obvious pulsation has been detected. In the TeV energy band, 507-day observations by HAWC find no $\gamma$-ray emission \citep{2017ApJ...843...40A,2020A&A...643A..28D}, while in LHAASO's recent results, a very large-scale extended emission detected by WCDA is spatially coincident with the extended radio/GeV emission, however, in the higher energy band, the source size reported by KM2A is much smaller centered around the point like soft $\gamma$-ray source 4FGL J0426.5+5434 \citep{2023arXiv230517030C}.

In the present work, we carry out a detailed analysis of the $\gamma$-ray emission toward G150.3+4.5 region using 14 years \emph{Fermi}-LAT data in the energy range of 100 MeV - 1 TeV in Sect. \ref{sec:2}. In Sect. \ref{sec:co}, the CO (J = 1-0) observation results of molecular clouds in this region is presented to explore the origin of the $\gamma$-ray emission. 
In Sect. \ref{sec:4}, we discuss the potential interpretation of the $\gamma$-ray emission based on multi-wavelength observational constraints. Lastly, Sect. \ref{sec:5} offers our conclusions.

\section{\emph{Fermi}-LAT data reduction}\label{sec:2}

   \textit{Fermi}-LAT is a pair-conversion instrument that is sensitive to $\gamma$-rays from 20 MeV to several hundreds of GeV (an overall description can be seen in\citep{Atwood2009}). In the following analysis, the newest Pass 8 data are collected from August 4, 2008 (mission elapsed time 239557418) to August 4, 2022 (mission elapsed time 681264005) to study the GeV emission around SNR G150.3+4.5 labelled in Data Release 3 of fourth \emph{Fermi}-LAT source catalog 4FGL-DR3;\citep{2020ApJS..247...33A,2022ApJS..260...53A}. We also select the data with "Source" event class ``P8R3$\_$SOURCE'' (evclass=128) and event type FRONT + BACK (evtype=3) with the standard data quality selection criteria $\tt (DATA\_QUAL > 0)  \&\& (LAT\_CONFIG $ == 1), and exclude the zenith angle greater than 90$\degr$ to avoid the earth limb contamination. To derive a better point-spread function, the energies of photon are cut between 2 GeV and 1 TeV to do further morphological analysis, and events with an energy between 100 MeV and 1 TeV are selected to do a more detailed spectral analysis of each spatial component. Events within a $14\degr\times14\degr$ region of interest (ROI) centered at the position of G150.3+4.5 are considered for the binned maximum likelihood analysis \citep{mattox1996likelihood} together with the instrument response functions (IRF) ``P8R3\_SOURCE\_V3''. The Galactic/isotropic diffuse background models (IEM, $\tt gll\_iem\_v07.fits$)/($\tt iso\_P8R3\_SOURCE\_V3\_v1.txt$ ) are adopted, and all sources listed in the 4FGL-DR3 catalog are included in the background model except for those within $7\degr$ from the center of ROI, whose normalizations and spectral indices are set free, and the normalizations of galactic/extra-galactic diffuse emission are also set free with the software make4FGLxml.py\footnote{\url{ http://fermi.gsfc.nasa.gov/ssc/data/analysis/user}}. The maximum likelihood test statistic (TS) is used to estimate the significance of $\gamma$-ray sources, where TS $= 2 (\ln\mathcal{L}_{1}-\ln\mathcal{L}_{0})$, and $\mathcal{L}_{1}$ and $\mathcal{L}_{0}$ are likelihood values for the background with and without target source (null hypothesis), respectively.

\subsection{Morphological analyses}

\begin{figure*}
    \centering
    \includegraphics[trim={0 0.cm 0 0}, clip,width=0.46\textwidth]{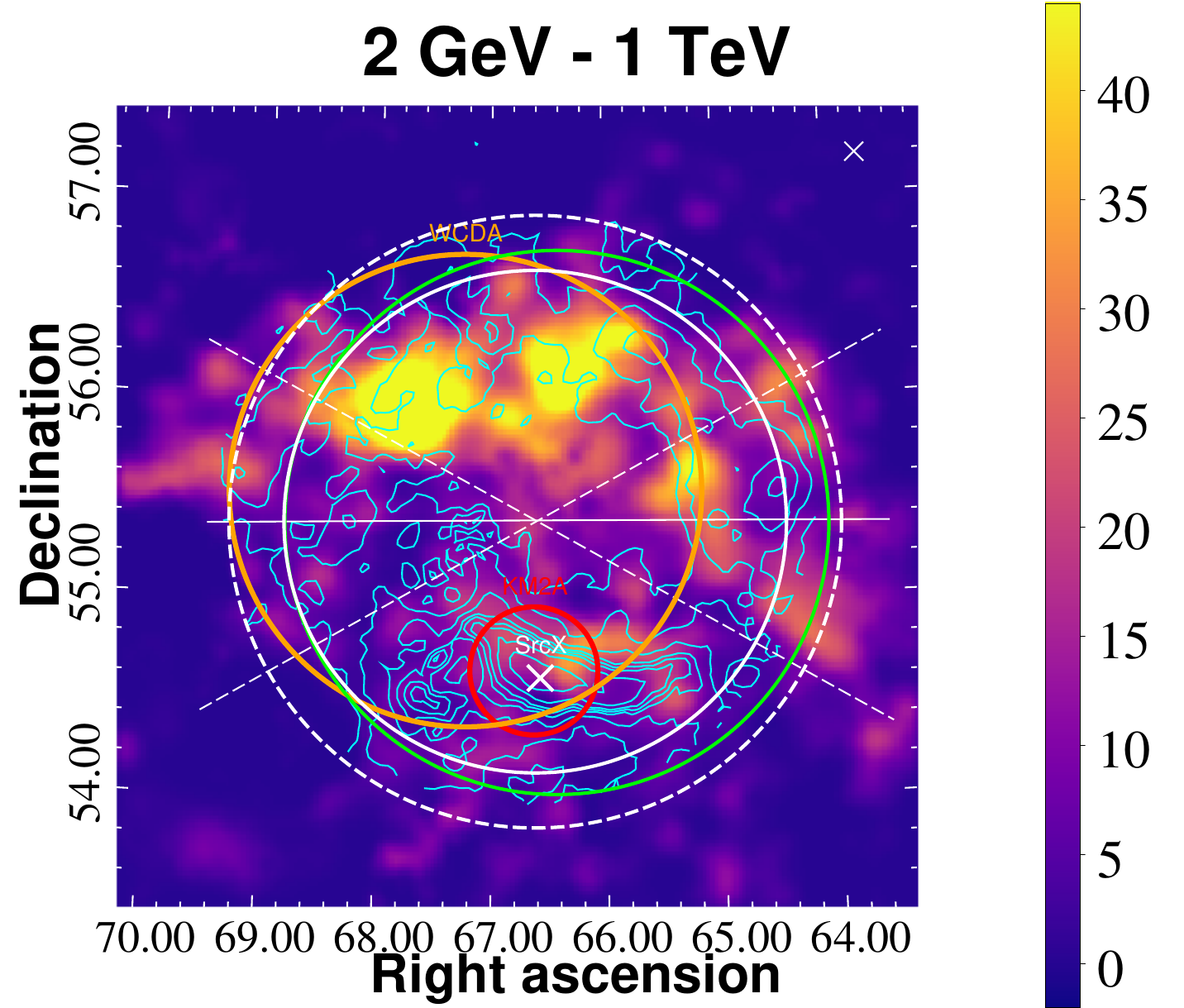}
    \includegraphics[trim={0 0.cm 0 0}, clip,width=0.46\textwidth]{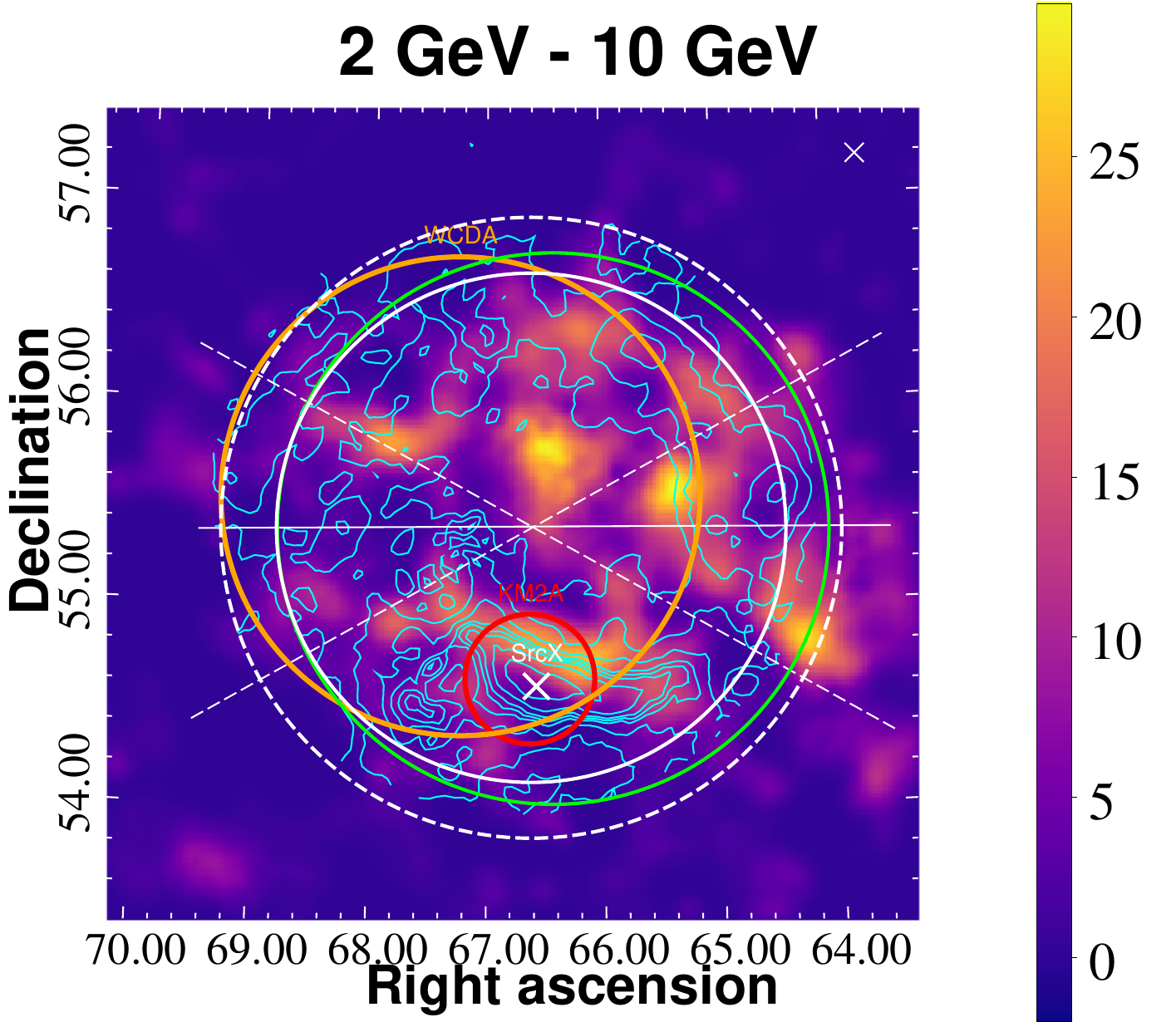}\\
    \includegraphics[trim={0 0.cm 0 0}, clip,width=0.46\textwidth]{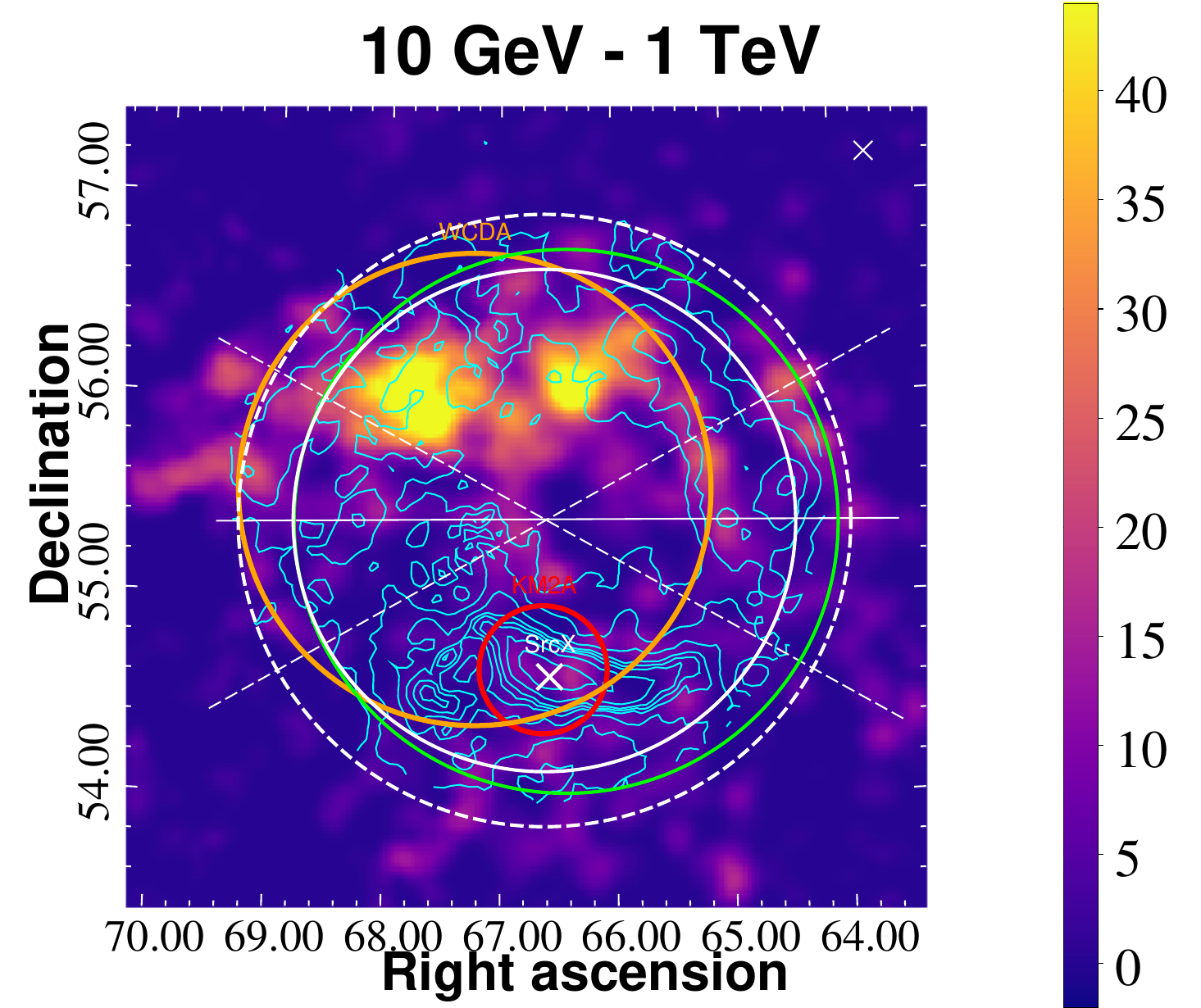}
    \includegraphics[trim={0 0.cm 0 0}, clip,width=0.46\textwidth]{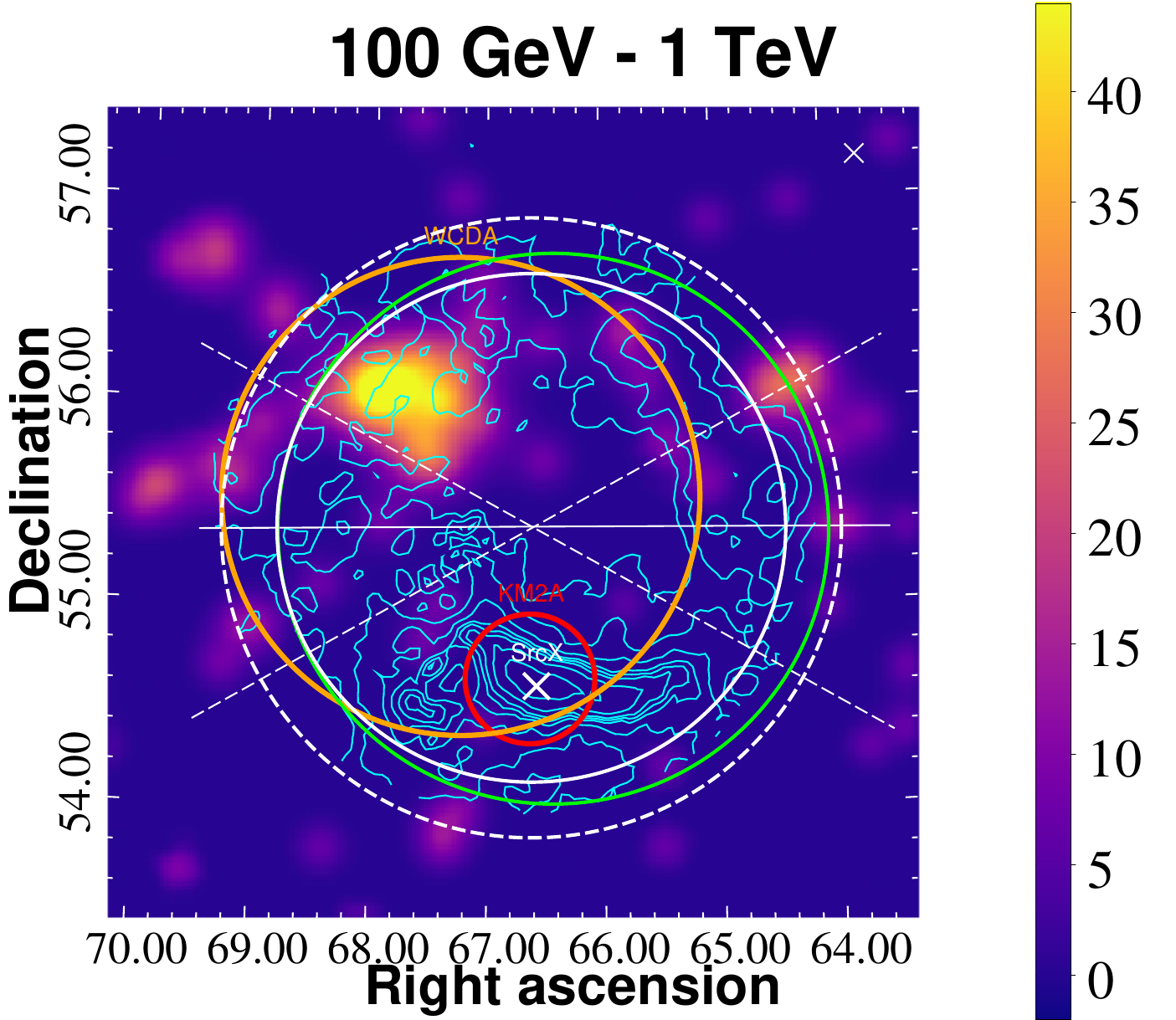}
    \caption{TS maps in the vicinity of SNR G150.3+4.5 observed with the \emph{Fermi}-LAT. All figures are generated with square bins of 0.02$\degr$. The energy ranges are indicated above the panels.
    The green circle represents the best-fit 2-D Gaussian template from \citet{2020A&A...643A..28D}. The white solid and dashed circles show the  $R_{68}$ and radius of the uniform disk model in this work, which is separated by the white horizontal line into two half-spheres: NorthLobe and SouthLobe (models 8 $\&$ 11 $\&$ 12). The white dashed line corresponds to the +30$\degr$ and -30$\degr$ scenarios (models 9 $\&$ 10). The unnamed white cross in the upper right corner represents a background point source from 4FGL-DR3\citep{2020ApJS..247...33A,2022ApJS..260...53A}, and SrcX represents the point source 4FGL J0426.5+5434, that has been treated as a distinct background point source. The cyan contours are extracted from Urumqi $\lambda$6cm results\citep{2014A&A...567A..59G}, showing the complete shell structure. The red and orange circles are $\rm R_{39}$($\sigma$) size for the LHAASO KM2A and WCDA sources\citep{2023arXiv230517030C}, respectively.}
    \label{fig:1}
\end{figure*}

The GeV emission in the SNR G150.3+4.5 region is described by an extended source 4FGL J0425.6+5522e plus a point-like source 4FGL J0426.5+5434 in 4FGL-DR3 \citep{2022ApJS..260...53A}. The former is suggested to be associated with SNR G150.3+4.5. To avoid contamination from the point source 4FGL J0426.5+5434, we use the gtfindsrc command to find its best fit location (R.A. = $66.585^{\circ}\!\pm0.017^{\circ}\!$, Dec. = $54.594^{\circ}\!\pm0.015^{\circ}\!$ with the 68$\%$ error radii r$_{\rm 68}$ = $0.025^{\circ}\!$). Here, the 68$\%$ error radii correspond to the radii of the 68\% confidence level contours for point-like sources. This measurement is almost the same as the position previously recorded in \citet{2020A&A...643A..28D,2022ApJS..260...53A}. To search for the energy dependent morphology of the GeV emission towards the SNR G150.3+4.5 region, all TS maps shown in Fig. \ref{fig:1} are generated by only considering background fitting without including G150.3+4.5. The $\gamma$-ray emission in different energy bands are shown in Fig. \ref{fig:1}, and for convenience, we indicate 4FGL J0426.5+5434 as SrcX. Above 2 GeV (top left), there are highly asymmetric diffuse $\gamma$-ray emission with several bright excesses mainly distributed in the north region, and some of them have good spatial correspondence with radio intensity. Above 10 GeV (bottom left), the emission located in the southern region gradually disappears, while some parts in the North are still bright. Thus we additionally generated the TS map in 2 - 10 GeV band (top right) that shows a more or less uniform brightness profile. Above 100 GeV (bottom right), the emission located in the southern part disappears completely, while the northern part is still significant. 

To further determine the best spatial template of the GeV emission in this region, we tested several spatial templates under the assumption of power-law spectral type with both the normalization and spectral index set free. We first added one point source (model 1) in the strongest excess location and used the gtfindsrc command to find the best location, recorded as (R.A. = $67.7207^{\circ}\!\pm0.03^{\circ}\!$, Dec. = $55.9180^{\circ}\!\pm0.05^{\circ}\!$,  r$_{\rm 68}$ = $0^{\circ}\!.11$). The TS value derived is only 81 with 4 degrees of freedom. Then, we added another point source (model 2) into another brightest excess location and we repeated the gtfindsrc command to search for the best-fit location, recorded as (R.A. = $66.4013^{\circ}\!\pm0.02^{\circ}\!$, Dec. = $56.0637^{\circ}\!\pm0.03^{\circ}\!$,  r$_{\rm 68}$ = $0^{\circ}\!.08$). The TS value derived is 147 with 8 degrees of freedom. Considering that there are still several multiple point-like excesses emerging in the 2 - 10 GeV energy band, we also tested the multi-point sources hypothesis (model 3). In this case, we gradually added point-like sources and repeated the gtfindsrc command to search the best location under power-law spectral shape assumption, until all excess peaks can be removed cleanly and the TS value reaches the maximum. 7 point-like sources are found in total. The TS value derived is 296 with 28 degrees of freedom. Then we adopted Fermipy tool \citep{2017ICRC...35..824W} to quantitatively evaluate the extension and location of the extended source scenario, like a single uniform disk (model 4) and a 2-D Gaussian (model 5) template. Here, r$_{\rm 68}$ is the 68\% containment radius and r$_{\rm 68_{Disk}}$ = 0.82$\sigma$, r$_{\rm 68_{Gaussian}}$ = 1.51$\sigma$ as suggested by \citet{2012ApJ...756....5L}, and the best-fit results are R.A. = $66.6151^{\circ}\!\pm0.02^{\circ}\!$, Dec. = $55.3755^{\circ}\!\pm0.02^{\circ}\!$,  r$_{\rm 68_{Disk}}$ = $1.255^{\circ}\!\pm0.03^{\circ}\!, \sigma$ = $1.530^{\circ}\!\pm0.04^{\circ}\!$ and R.A. = $66.4258^{\circ}\!\pm0.05^{\circ}\!$, Dec. = $55.1577^{\circ}\!\pm0.04^{\circ}\!$,  r$_{\rm 68_{Gaussian}}$ = $1.377^{\circ}\!\pm0.05^{\circ}\!, \sigma $= $0.912^{\circ}\!\pm0.03^{\circ}\!$, similar to the results from \citet{2020A&A...643A..28D}. In addition, we produced a series of ring templates (model 6) at the center of the uniform disk template with various inner radius from $0^{\circ}\!.1$ to $0^{\circ}\!.8$ in steps of $0^{\circ}\!.05$. The outer radius is fixed to the uniform disk template. We found that the TS values and likelihood values did not increase significantly, and the highest TS value case is shown in Table \ref{tab:1}. To search for the spatial correlation between GeV morphology and radio intensity, we also test Urumqi $\lambda$6cm map (model 7) as a template. However, its TS value is still low. To further substantiate the potential spectral variations between the northern and the southern region depicted in Fig. \ref{fig:1}, we divide the uniform disk template into two half-disks (model 8) along a western/eastern line in celestial coordinates. Then we further conducted additional tests on two half-disk templates with different slope lines, which are not horizontal but represent +30 and -30 degrees (model 9 and model 10), respectively. However, both of them had lower TS value performances compared with model 8. To further determine whether the extended GeV emission is correlated with gas distributions, we tested a spatial template for H$_{\rm2}$, which is produced from the carbon monoxide (CO) composite survey described in Sect. \ref{sec:co}. Similarly, we also divide the potential H$_{\rm2}$ template along the western/eastern line in celestial coordinates, recorded as model 11 and model 12, corresponding to different gas velocity intervals. It is important to note that, in the last two models, the gas distribution template was only applied to the southern half-disk region since molecular clouds are only distributed in the south (shown as Fig. \ref{fig:4}, details are discussed in the following Sect. \ref{sec:co}), the northern region is still described as a uniform half-disk template as in the model 8.

\begin{figure*}
    \centering
    \includegraphics[trim={0 0.cm 0 0}, clip,width=0.33\textwidth]{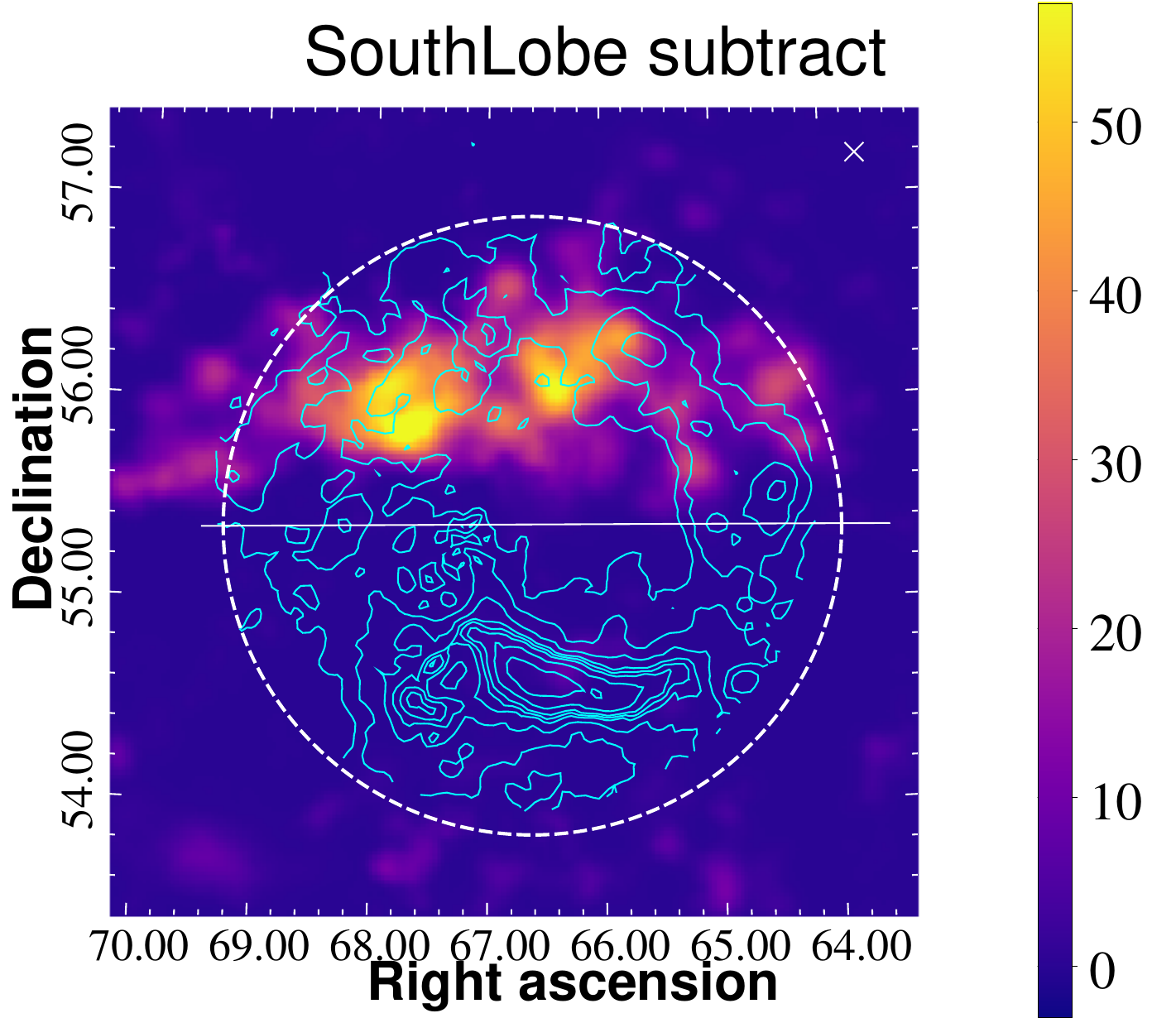}
    \includegraphics[trim={0 0.cm 0 0}, clip,width=0.325\textwidth]{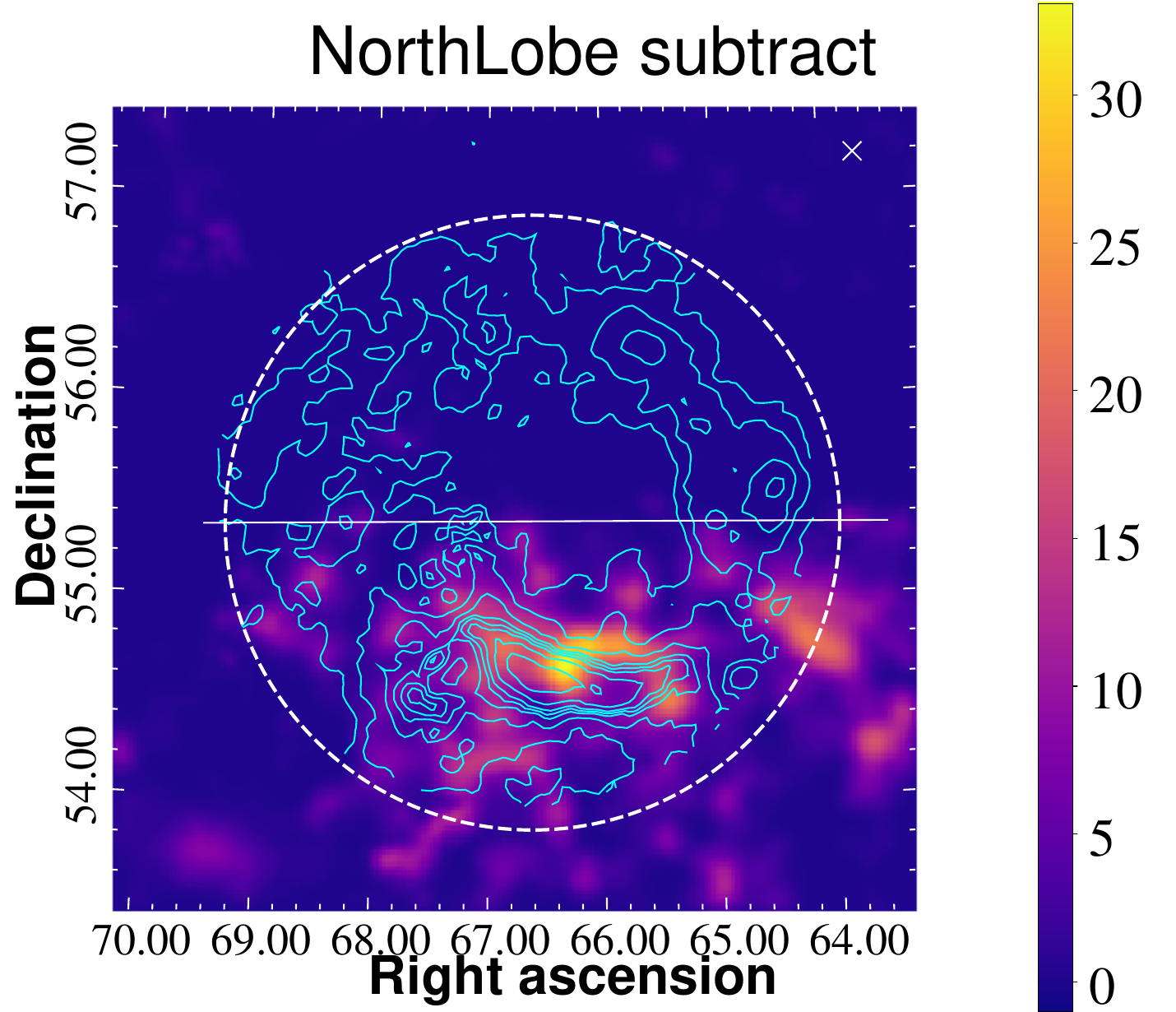}
    \includegraphics[trim={0 0.cm 0 0}, clip,width=0.325\textwidth]{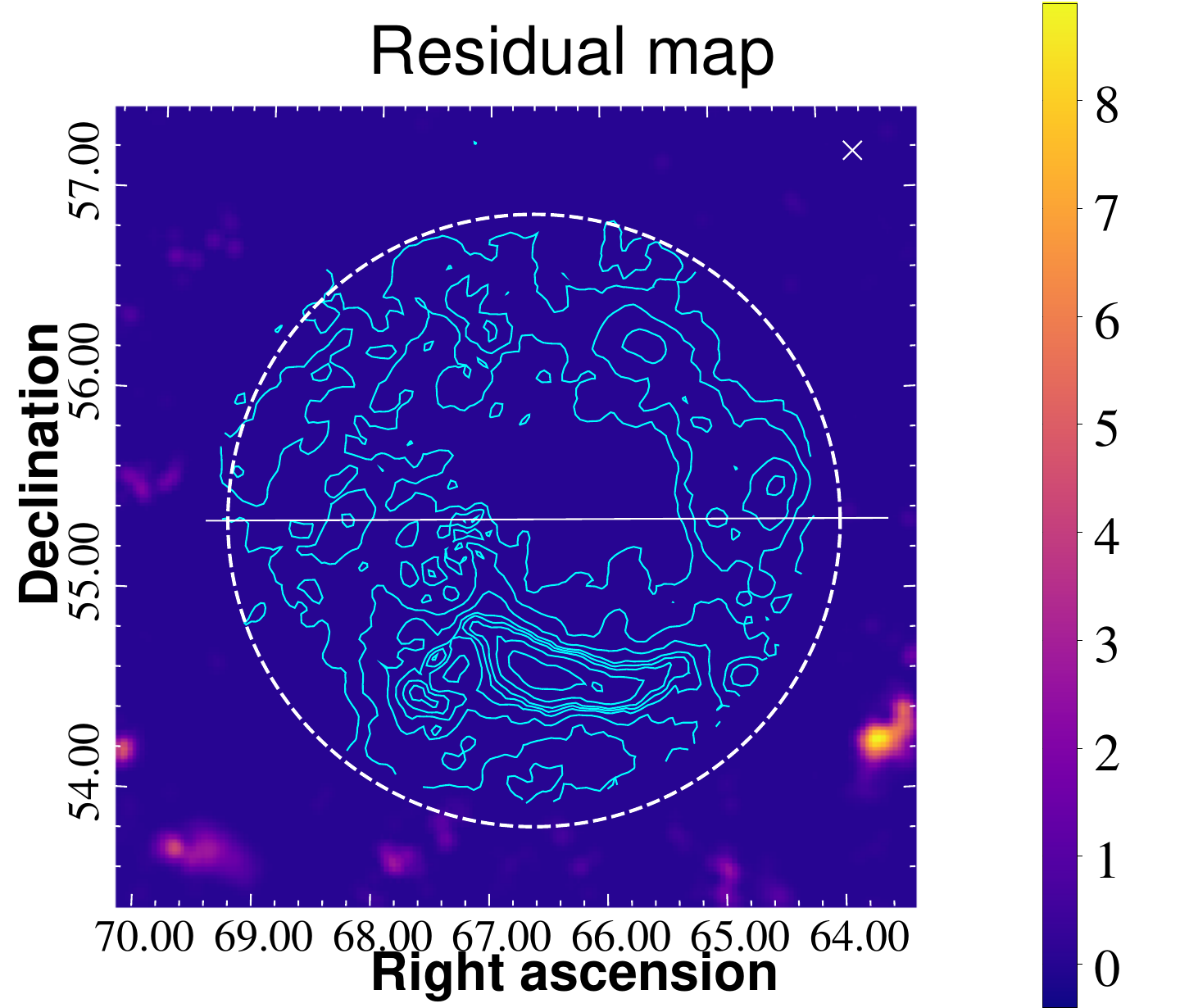}
    \caption{Residual maps above 2 GeV, obtained by subtracting southern/northern/both lobe, all labels match those in Fig. \ref{fig:1}.}
    \label{fig:residual}
\end{figure*}

The $\rm TS_{ext}$ is defined as $\rm {TS_{ext}}$=$2(\ln\mathcal{L}_{\rm ext}-\ln\mathcal{L}_{\rm ps})$,
where $\mathcal{L}_{\rm ext}$/$\mathcal{L}_{\rm ps}$ represents the maximum likelihood value for the extend/point-like template, considering only 1 free parameter added by the extended template, and the extension significance is approximately given by $\sqrt{\rm TS_{ext}}$ $\sigma$. When $\rm TS_{ext}$ $>$ 16 \citep{2012ApJ...756....5L}, the extended source hypothesis is valid. In model 4 (Disk), $\rm TS_{ext}$ is calculated as 542, which rejects the point-like source hypothesis at $23.3\sigma$. Model 5 (Two-dimensional Gaussian) shown a marginal improvement, $\rm TS_{ext}$ = 563 (correspond to $23.7\sigma)$. While the results from model 6/7 (ring and radio intensity map) are also similar, and the TS values are less than previous two templates. The two half-disk template (model 8) shows a better performance both in TS (marginal improvement $\sim$ 2.8$\sigma$) and AIC value, which is defined as AIC = $2k - 2\ln\mathcal{L}$ and describe in \citet{1974AIC}, where $k$ is the number of degrees of freedom of the model and $\mathcal{L}$ is the likelihood value. Additionally, the gas distribution template in different velocity ranges (model 11 $\&$ model 12) have similar improvements $\sim$ 3.7$\sigma$. Model 11 has the lowest value of AIC and the highest TS value, and is selected to carry out the subsequent spectrum analysis. We also analysed the energy spectrum of the Disk template (model 4) for comparison. Hereafter, we will refer to the northern/southern lobes derived from H$_{\rm2}$ template (model 11) as $\rm{NorthLobe}$ and $\rm{SouthLobe}$, respectively, and the residual maps obtained after subtracting either the southern, northern or both lobes independently are shown in the Fig. \ref{fig:residual}. The above morphology analysis results are summarized in Table \ref{tab:1}. Furthermore, we tested the energy dependent extension of SNR G150.3+4.5 in each energy band, and the results suggest that the extension size does not shrink visibly. No significant energy-dependent morphology is found. The results are summarized in Table \ref{tab:energy-dependent}.

\begin{table*}  
    \caption{Spatial models tested for the GeV $\gamma$-ray emission above 2 GeV(SrcX subtracted )} \label{tab:1}
    \centering
    \begin{tabular}{lcccccc}
    \hline
Morphology($>$2GeV) & TS & TS$_{\rm ext}$ & Best$-$fit extension($\degr$)&$\sigma$  ($\degr$)&Ndf$^{\,\,\text{a}}$&$\Delta${AIC}$^{\,\,\text{b}}$\\
\hline
Model 1 (Single point)&81&$-$&$-$&$-$&4&0 \\
Model 2 (Two point)&147&$-$&$-$&$-$&8&-62 \\
Model 3 (Multiple point)&296&$-$&$-$&$-$&28&-167 \\
Model 4 (Disk)&623&542&R$_{68}$=1.255$\pm 0.03$ &  $\sigma$=1.530$\pm0.04$ & 5&-540 \\
Model 5 (Gaussian)&644&563&R$_{68}$=1.377$\pm 0.05$ & $\sigma$=0.912$\pm 0.03$ &5&-561 \\
Model 6 (Ring template)$^{\,\,\text{c}}$&359&278&R$_{\rm{in(out)}}$=0.15$\degr$(1.53$\degr$)&$-$&6&-274 \\
Model 7 (Urumqi $\lambda$6cm map)&351&270&$-$&$-$&2&-274 \\
Model 8 (Two half-lobes horizon)&652&571&$-$&$-$&7&-565 \\
Model 9 (Two half-lobes +30$\degr$)&651&570&$-$&$-$&7&-564\\
Model 10 (Two half-lobes -30$\degr$)&649&568&$-$&$-$&7&-562\\
Model 11 (H$_{2}$ map 1)$^{\,\,\text{d}}$&658&577&$-$&$-$&7&-571\\
Model 12 (H$_{2}$ map 2)$^{\,\,\text{e}}$&657&576&$-$&$-$&7&-570\\
\hline
    \label{table:1}
    \end{tabular}\\
{{\bf Notes.} $^{(a)}$ Degrees of freedom. $^{(b)}$ Calculated with respect to model 1. $^{(c)}$ Best-fit ring inner radius in steps of 0.05\degr. $^{(d)}$ Gas velocity interval in [-9.8,+2.4] km s$^{-1}$. $^{(e)}$ Gas velocity interval in [-9.8,-3.8] km s$^{-1}$.}
\end{table*}



\begin{table*}[h]
    \centering
    \caption{The results of energy-dependent extension measurements of SNR G150.3+4.5 with 2-D Gaussian template.}
    \begin{tabular}{cccccccc}
    \hline\hline
    Energy range     & Best-fit position                                                                                    & Extension $r_{68}$                                                        & TS &TS$_{\rm ext}$        \\
                                              & (R.A., decl.)                                                                                        &                                                                           &                     \\ \hline
    2 GeV - 10 GeV  & ($66\overset{\circ}{.}43\pm0\overset{\circ}{.}06, 55\overset{\circ}{.}16\pm0\overset{\circ}{.}05$) & $1\overset{\circ}{.}31_{-0\overset{\circ}{.}07}^{+0\overset{\circ}{.}06}$ & 168 & 137  \\                 
     10 GeV - 100 GeV   & ($66\overset{\circ}{.}78\pm0\overset{\circ}{.}05, 55\overset{\circ}{.}49\pm0\overset{\circ}{.}05$) & $1\overset{\circ}{.}38_{-0\overset{\circ}{.}05}^{+0\overset{\circ}{.}06}$ & 381 & 310 \\
    100 GeV - 1 TeV   & ($66\overset{\circ}{.}82\pm0\overset{\circ}{.}04, 55\overset{\circ}{.}61\pm0\overset{\circ}{.}05$) & $1\overset{\circ}{.}17_{-0\overset{\circ}{.}05}^{+0\overset{\circ}{.}07}$ & 94 & 42\\ \hline
    \end{tabular} 
    \label{tab:energy-dependent}
\end{table*}

\subsection{Energy spectrum}
\begin{figure*}
    \centering
    \includegraphics[trim={0 0.cm 0 0}, clip, width=0.325\textwidth]{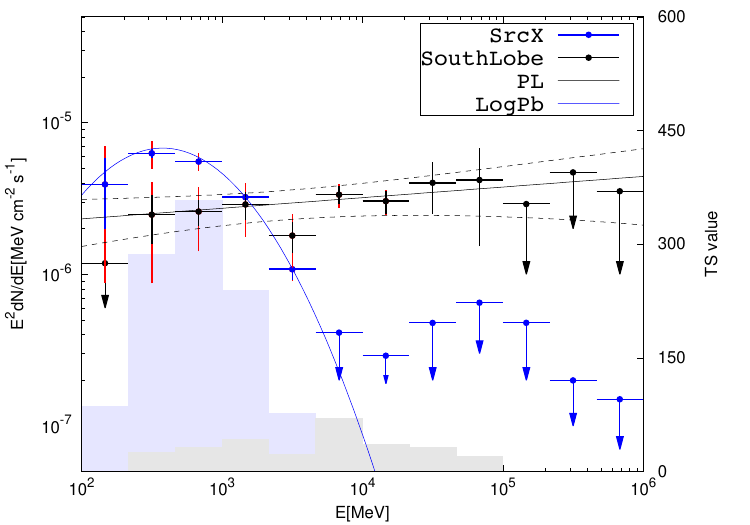}
    \includegraphics[trim={0 0.cm 0 0}, clip, width=0.325\textwidth]{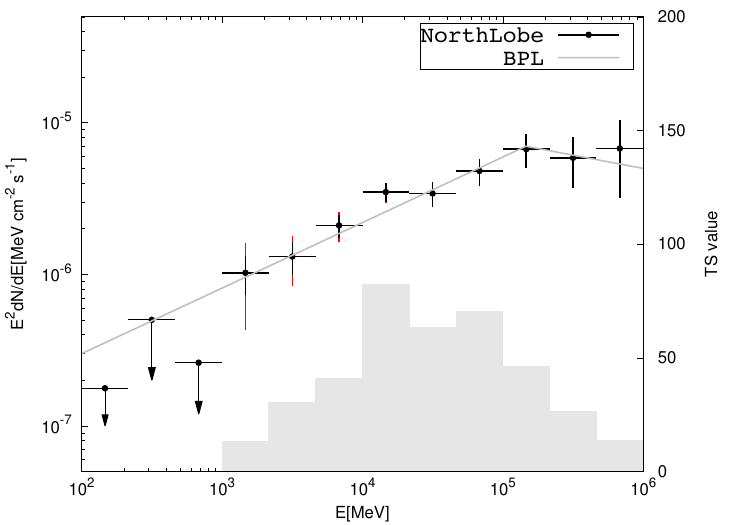}
    \includegraphics[trim={0 0.cm 0 0}, clip, width=0.325\textwidth]{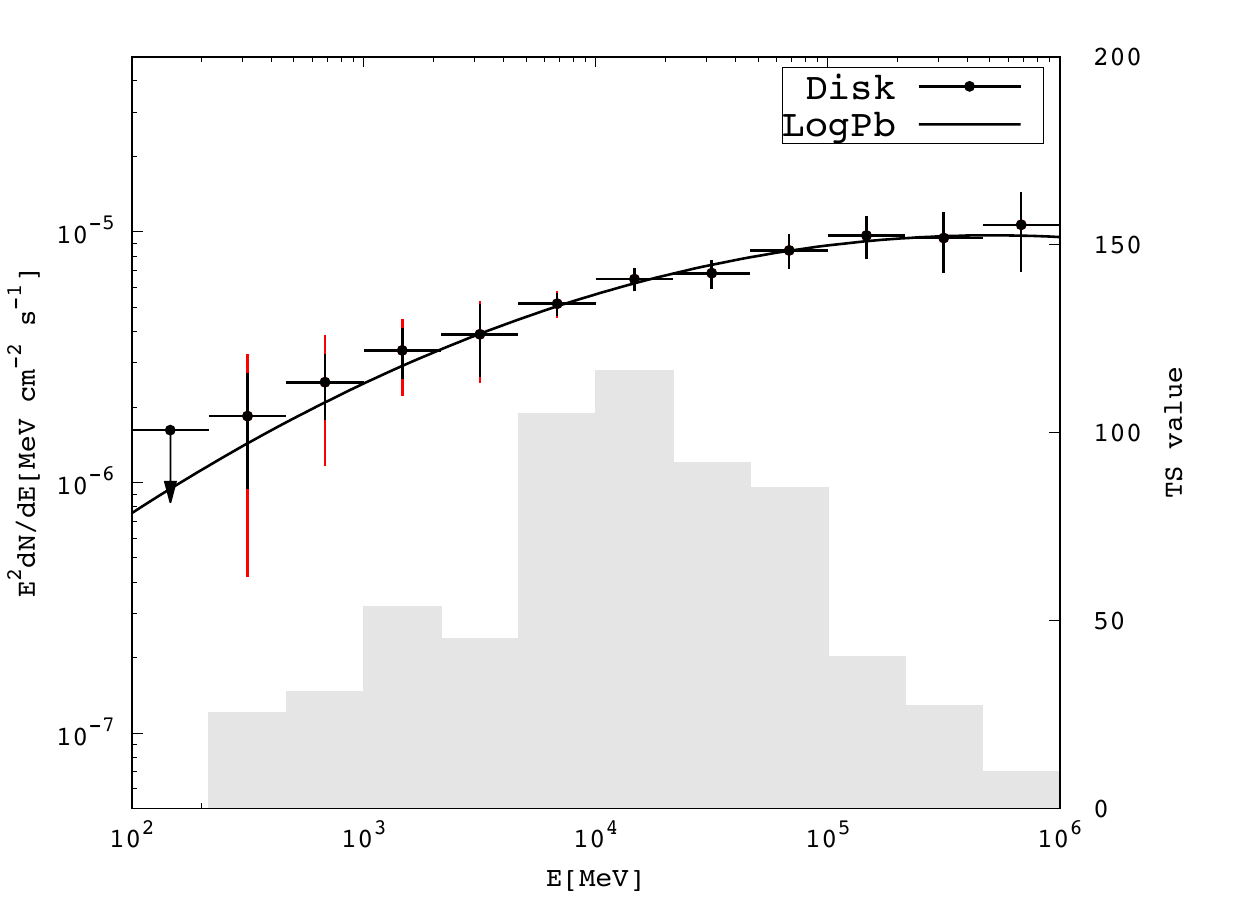}
    \caption{SED and TS values (shaded areas) of the SouthLobe/SrcX (panel left), NorthLobe (panel middle) and Disk (panel right) as measured by \emph{Fermi}-LAT in the energy range from 0.1 to 1000 GeV. Upper limit are calculated at 95\% confidence level using a Bayesian method for points with TS values less than 5. The statistical errors are shown in black/blue error bars, while the sum of statistical and systematic errors calculated by $\sigma=\sqrt{\sigma_{\rm stat}^2+\sigma_{\rm sys}^2}$ are marked by red error bars. The Disk and SrcX can be described with a log-parabola spectrum. The NorthLobe is best-fitted with a broken power-law spectrum. The SouthLobe is compatible with a single power-law and the black dash lines show the 1 $\sigma$ statistic error uncertainty.}
    \label{fig:2}
\end{figure*}

    Given the best-fit spatial template model 11 determined above, we perform $\gamma$-ray spectral analyses for each components using larger amount of events from 100 MeV to 1 TeV. Considering previous results shown in \citet{2020A&A...643A..28D}, we adopted logparabola (LogPb; dN/dE $\propto$ E$^{\rm -(\alpha+\beta log(E/E_{\rm b}))}$) spectra directly for Disk (Model 4) and SrcX. For NorthLobe and SouthLobe (model 11), we tested different spectral types and the spectral indices show a significant difference in both simple power-law (PL; dN/dE $\propto$ E$^{-\alpha}$), and broken power-law (BPL \footnote{$dN/dE\propto
        \begin{cases}
        \left(\frac{E}{E_b}\right)^{-\Gamma_1}, & E < E_{b} \\
        \left(\frac{E}{E_b}\right)^{-\Gamma_2}, & E > E_{b} \\
        \end{cases}$}
) assumptions as shown in Table \ref{table:2}. In this table, we calculate the estimated systematic errors due to the Galactic diffuse emission by changing the normalization of the best-fit Galactic diffuse model artificially by $\pm$6$\%$ as suggested in \citet{2010ApJ...714..927A}. The sums of statistical and systematic errors are calculated by $\sigma=\sqrt{\sigma_{\rm stat}^2+\sigma_{\rm sys}^2}$, which are shown as the red error bars in Fig. \ref{fig:2}. For the NorthLobe, compared with a single PL model, the BPL model has a notable improvement of the fitting quality, which could be quantify as  $\rm{TS_{curve}}$, defined as $\rm{TS_{curve}}$=$2(\ln\mathcal{L}_{\rm BPL}-\ln\mathcal{L}_{\rm PL})$\citep{abdollahi2020a}. The obtained value of 24 corresponds to a significance level of $\sim$ 4.9 $\sigma$ with only one additional free parameter. Thus we conclude that there is an energy spectral break at $\sim$146 GeV in the NorthLobe spectrum. For the SouthLobe, there is almost no variation in the TS value between PL and BPL cases, thus the SouthLobe can be described by a PL. Furthermore, to derive the spectral energy distributions (SEDs) of both sources, we divided the events in the 100 MeV - 1 TeV energy range into twelve logarithmically equal intervals and repeated the same likelihood fitting analysis for each interval. The normalizations of all sources are left free, and the spectral indices and energy break are fixed to their best-fit value. For bins with TS values less than 5.0, we give upper limits calculated with 95\% confidence level using a Bayesian method \citep{helene1983}. The SED of SrcX  exhibits a break around 1 GeV although the flux is insignificant above 10 GeV. Thus the $\gamma$-ray excess from SrcX is mainly in the lower GeV band implying a distinct spectral component for the extended TeV emission detected by LHAASO-KM2A. 
To further test the flux stability for SrcX, we analyzed its light curve (LC) in the following part.

\begin{table*}[h]  
    \caption{Spectral fit parameters between 0.1-1000 GeV} \label{tab:2}
    \centering
    \begin{tabular}{lcccccc}
    \hline
Spatial model& Spectral type&\textbf{$\Gamma$}1($\alpha$)$^{\,\,\text{a}}$&\textbf{$\Gamma$}2($\beta$)$^{\,\,\text{b}}$&E$_b$(GeV)&Photon flux(photon cm$^{-2}$ s$^{-1}$)&TS \\
\hline
Disk             &LogPb&1.72$\pm$0.08$\pm$0.13&0.035$\pm$0.005$\pm$0.008&9.37&(6.38$\pm$0.35)$\times$ 10$^{-8}$&623 \\
\hline
SrcX &LogPb&2.44$\pm$0.06$\pm$0.08&0.40$\pm$0.07$\pm$0.09&0.65&(9.17$\pm$0.62)$\times$10$^{-8}$&949 \\
\hline
\multirow{2}*{NorthLobe}&PL&1.56$\pm$0.11$\pm$0.17&$-$&$-$&(2.98$\pm$0.11)$\times$10$^{-8}$&366 \\
~&BPL& 1.54$\pm$0.04$\pm$0.07 & $2.28\pm0.08$ $\pm$0.12 & 146$\pm$11 & (2.94$\pm$0.12)$\times$10$^{-8}$ & 390 \\
\hline
\multirow{2}*{SouthLobe}&PL&1.95$\pm$0.07$\pm$0.09&$-$&$-$&(1.54$\pm$0.51)$\times$10$^{-8}$&262 \\
~&BPL& 1.96$\pm$0.08$\pm$0.08 & 2.03$\pm$0.11$\pm$0.10 & 8.1$\pm$2.2 & (1.55$\pm$0.53)$\times$10$^{-8}$ & 263 \\

\hline
    \label{table:2}
    \end{tabular}
{{\bf Notes.} $^{(a)}$$^{(b)}$ Best-fit value $\pm$ Statistical error $\pm$ Systematic error.}
\end{table*}

\subsection{Flux variability of SrcX}\label{sec:flux}

\begin{figure}
    \centering
    \includegraphics[trim={0 0.cm 0 0}, clip, width=0.35\textwidth]{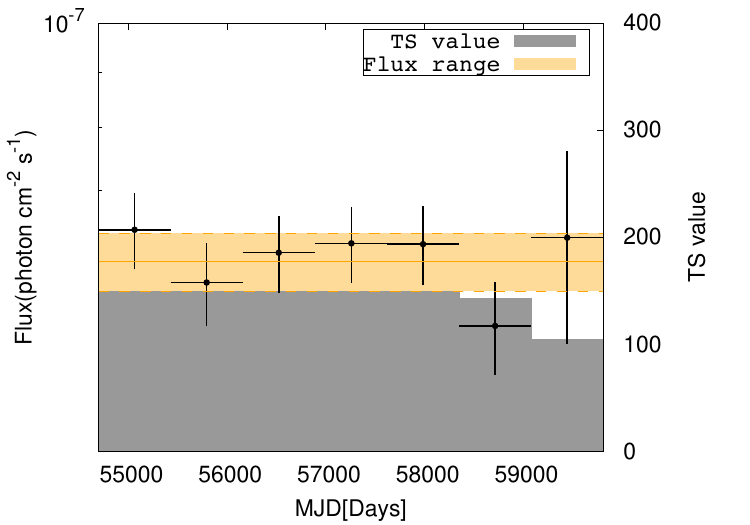}
    \caption{Light curves of SrcX from 0.1 to 10 GeV. The mean flux and its 1 $\sigma$ statistic uncertainty are shown as the orange horizontal solid and dashed lines, respectively. The grey histogram represents the TS value for each time bin.}
    \label{fig:3}
\end{figure}

Considering the complex environment around SrcX, some potential unidentified counterpart that might contributed to $\gamma$-ray excess in its 95$\%$ error circle through the SIMBAD \footnote{\url{ http://simbad.u-strasbg.fr/simbad/sim-fbasic}} database, like Galaxy LEDA 2471506\citep{2003A&A...412...45P}, Nebula DSH J0426.0+5433\citep{2006A&A...447..921K} and variable star ATO J066.5105+54.5272 detected by \emph{Gaia} \citep{2020yCat.1350....0G}, which leads its $\gamma$-ray excess puzzling, thus it is essential to explore its flux variability. The 14 yr data with energies between 0.1 - 10 GeV were divide into seven time bins, respectively. For the timing analysis, all spectral indices of the sources included in the model were fixed to the best-fit values, only normalizations are allowed to vary, then we repeated the likelihood analysis in each time bin. For the time bin with a TS value smaller than 9, an upper limit with 95$\%$ confidence level was calculated. The derived light curve are shown in Fig. \ref{fig:3}. We also calculated the variability index as defined in \citet{2012ApJS..199...31N}, which under the assumption that with a value in the null hypothesis where the source flux is constant across the full time period, and the value under the alternate hypothesis where the flux in each bin is optimized as TS$_{\rm var}$ = $\sum_{i=1}^{N} 2 \times (\rm{Log}(\mathcal{L}_i(\mathcal{F}_i))-\rm{Log}(\mathcal{L}_i(\mathcal{F}_{\rm mean})))$, where $\mathcal{L}_i$ is the likelihood corresponding to bin $i$, $F_i$ is the best-fit flux for bin $i$, and $\mathcal F_{\rm mean}$ is the best-fit flux for the full time assuming a constant flux. For this LC, the source of TS$_{\rm var} \geq 16.81$ was considered to be a variable source with a 99$\%$ confidence level, given the TS$_{\rm var}$ is anticipated to be distributed as $\chi^2_{N-1}(\rm TS_{var})$ in the null case, we get TS$_{\rm var}$ = 6.8 for the flux variability using above energy band, which is very marginal, thus we suggests no significant variability for the $\gamma$-ray excess from SrcX. Nevertheless, we notice that the TS values in last two time bins has a insignificant decline. Considering that the data points still locate in 1 $\sigma$ uncertainty region, one may conclude that the photon flux is stable, and the evidence is too weak to draw a conclusion about flux variability, which can be interpreted as statistical fluctuations.

\section{CO observations}\label{sec:co}

\begin{figure*}
    \includegraphics[trim={0 0.cm 0 0}, clip, width=0.325\textwidth]{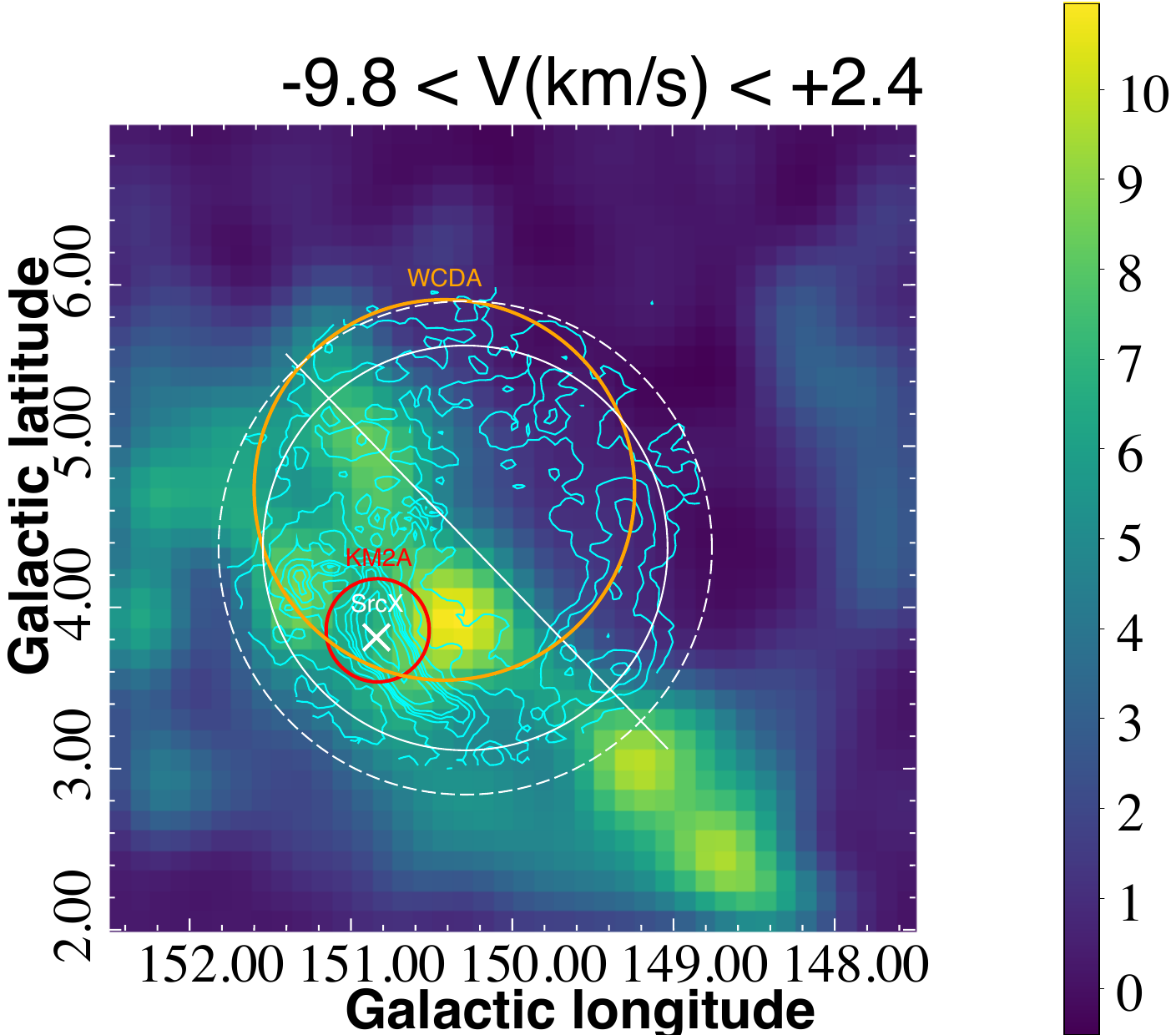}
    \includegraphics[trim={0 0.cm 0 0}, clip, width=0.325\textwidth]{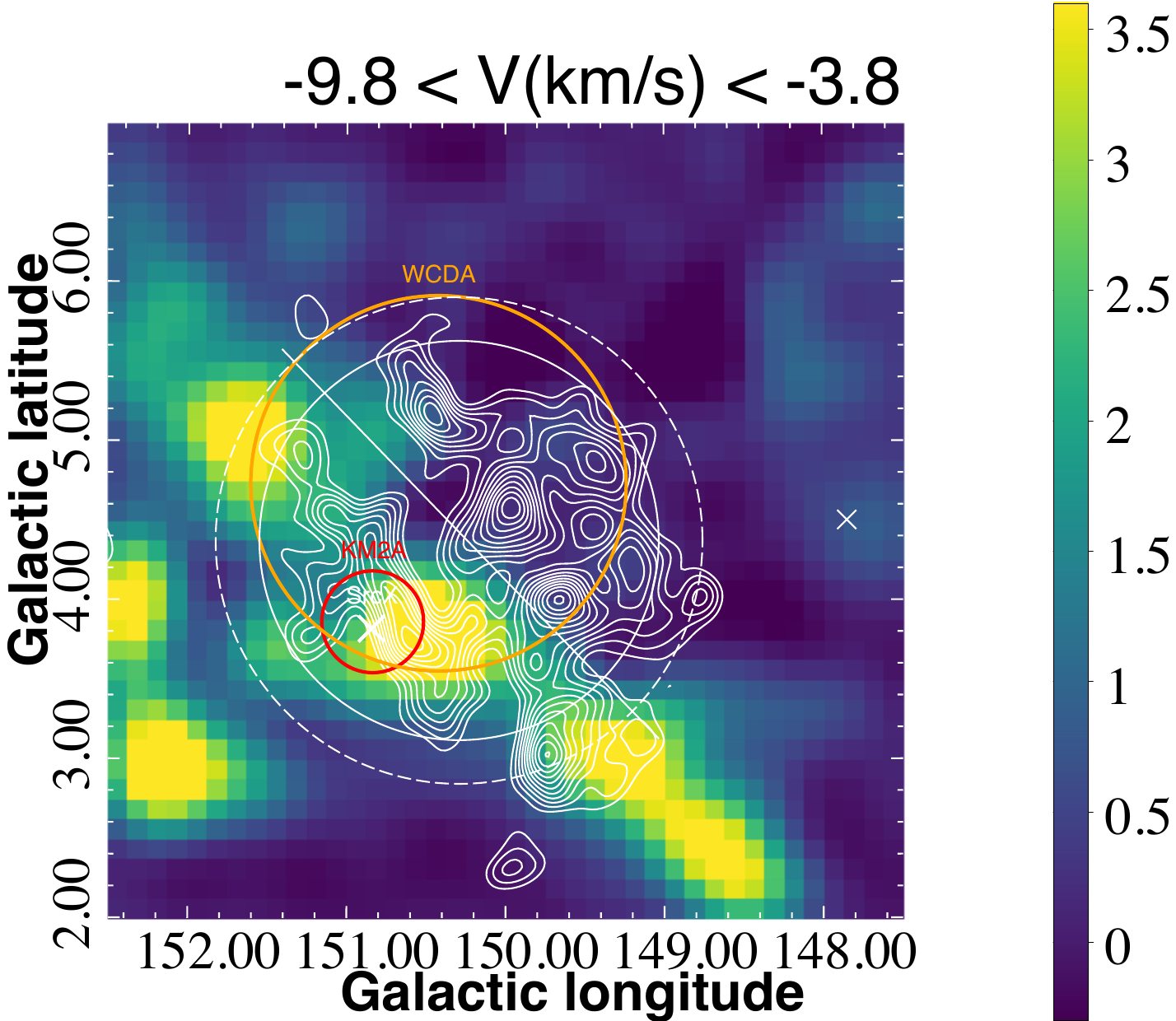}
    \centering
    \caption{Integrated CO (J = 1-0) emission intensity (K km s$^{-1}$) toward SNR G150.3+4.5 at 115 GHz using CO composite survey \citep{2001ApJ...547..792D}. Panel left : Integrated on the whole velocity range $\left[-9.8, +2.4\right]$ km s$^{-1}$, and  overlaid with the radio shell structure extracted from Urumqi $\lambda$6cm results\citep{2014A&A...567A..59G}. Panel right: Integrated on $\left[-9.8, -3.8\right]$ km s$^{-1}$ while overlaid with the GeV emission contour line generated from the top right panel in Fig. \ref{fig:1}. Other labels are same as above.}
\label{fig:4}
\end{figure*}

To track the distribution of H$_{\rm2}$ surrounding the SNR G150.3+4.5 region, we used the CO composite survey \citep{2001ApJ...547..792D} and the standard assumption of a linear relation between the velocity-integrated brightness temperature of CO (J = 1-0) line at 115 GHz (2.6mm) and the molecular hydrogen column density. The HI spectrum obtained from the Leiden/Argentine/Bonn survey by \citet{2005A&A...440..775K} reveals the presence of two velocity components within the range of $\left[-9.8, +2.4\right]$ km s$^{-1}$. This observation suggests that clouds in this study may have been disturbed by shocks, as discussed by \citet{2013A&A...557L..15C} and \citet{2017ApJ...845...48S}, which is further confirmed by recent PMO 13.7 m millimeter telescope\citep{2024A&A...686A.305F}, indicating that the signature of SNR associated with surrounding MCs is present in this region. The more precise velocity range ($\left[-9.8, +2.4\right]$ km s$^{-1}$) was considered as whole velocity range for analysis, including the velocity ranges $\left[-9.8, -3.8\right]$ km s$^{-1}$ discussed by \citet{2020A&A...643A..28D} for comparison, and depicted in the left/right panel of Fig. \ref{fig:4}. The whole velocity interval can be translated into a kinematic distance of $d$ = $0.55_{-0.55}^{+0.65}$ kpc \citep{2020A&A...643A..28D}. This finding is consistent with a kinematic distance of $d$ = 0.8 kpc and an age of 13 kyr reported by \citet{Zeng:2023uvu}. Given that the SNR is located inside the distance error range of MCs and their potential interactions, here we assume that the molecular clouds are associated with SNR G150.3+4.5 and we adopt the distance $d$ = 0.8 kpc in further calculations. Then $\theta$= $1^{\circ}\!.53$ corresponds to a physical radius of 21.4 pc. In order to evaluate the column density of H$_{2}$ in this region, a conversion factor $X_\mathrm{CO}$ = $2\times10^{20} \rm{cm^{-2}} \ \rm{K^{-1}} \ \rm{km^{-1}} \ \rm{s}$ is used \citep{bolatto2013,2001ApJ...547..792D}, and $N_\mathrm{H_2}$ can be derived as $N_\mathrm{H_2}$ = $X_\mathrm{CO} \times$ $W_\mathrm{CO}$. Thus the mass of the molecular complex can be calculated from $W_\mathrm{CO}$ as

\begin{equation}\label{eq:massco}
M = {\mu m_\mathrm{H}} D^2 \Delta\Omega_\mathrm{px} X_\mathrm{CO} {\sum_\mathrm{px}} W_\mathrm{CO} \propto N_\mathrm{H_2},
\end{equation}

where $\mu$ is equal to 2.8 if a relative helium abundance of 25$\%$ is assumed, $m_\mathrm{H}$ is the mass of the H nucleon, $N_\mathrm{H} = 2N_\mathrm{H_2}$ represents the column density of the hydrogen atom in each pixel, and $\Delta\Omega_\mathrm{px}$ corresponds to the solid angle subtended for each pixel in the map(square binning of 0.125$\degr$ per side). The term ${\sum_\mathrm{px}} W_\mathrm{CO}$ takes into account the binning in velocity of the data cube and is obtained by summing the map content for the pixels in the desired sky region and desired velocity range and scaled by the bin size in velocity.
Then we obtain the mass of the molecular cloud in the whole velocity range through its relation with $N_\mathrm{H_2}$, assuming a spherical geometry of the gas distribution, volume is given by $\rm{V_{Total}} = {{4\pi \over3}R^3}$ and $\rm{V_{NorthLobe/SouthLobe}} = 0.5\times \rm{V_{Total}}$, here R = d $\times$ $\theta$, the mass of gas is calculated to be $\rm{M_{SouthLobe}} = 3.3\times 10^{4}d_{0.8}^{2} \ M_{\odot}$, and the average $\rm H_2$ cubic density in this region is about $\rm{n_{SouthLobe}}$ = $\rm 66d_{0.8}^{-1} \ cm^{-3}$. For the north lobe, the mass of gas is estimated to be about $\rm{M_{NorthLobe}} = 4.7\times 10^{3}d_{0.8}^{2} \ M_{\odot}$, the average $\rm H_2$ cubic density is about $\rm{n_{NorthLobe}}$ = $\rm 9d_{0.8}^{-1} \, cm^{-3}$. For the Whole-disk region, the total mass of gas within $1.53\degr$ is estimated to be $\rm{M_{Total}} = 3.8\times 10^{4}d_{0.8}^{2} \ M_{\odot}$, average density is approximately $\rm{n_{Total}}$ = $\rm37d_{0.8}^{-1} \, cm^{-3}$, which is roughly consistent with the measurements from \citet{Zeng:2023uvu} and \citet{2024A&A...686A.305F}, while its differs noticeably from the $10^{-3}$ order estimated from X-ray measurements with ROSAT in \citet{2020A&A...643A..28D}. We suggest that the actual ambient density of about several tens $\rm cm^{-3}$ as derived from gas detector measurements will be more accurate. For the SrcX region, we estimate that the mass of gas within $0.1\degr$ around the central location of SrcX is calculated to be $\rm{M_{SrcX}} = 2.5\times 10^{2}d_{0.8}^{2} \ M_{\odot}$, corresponding to $\rm{n_{SrcX}}$ = $\rm 89d_{0.8}^{-1} \, cm^{-3}$, which indicates that there are some higher-density area in the southern region, but due to the resolution limit (0.125$\degr$) of CO composite survey data, we are currently unable to obtain the significantly fiber structure of gas distribution, and more high-resolution telescopes may be needed to observe this area.

As shown in Fig. \ref{fig:4}, there is good spatial correspondence between the gas distribution and the location of $\gamma$-ray emission in southern region, which might be the main contributor to SouthLobe. While in the northern region, there is almost no distribution of molecular cloud. To further investigate the origins of $\gamma$-ray emission in each part,  we model the $\gamma$-ray fluxes in the next section.

\section{A two-zone model}\label{sec:4}
\subsection{Hadronic emission dominated in the southern region}
The above {\em Fermi}-LAT data analysis shows that the GeV $\gamma$-ray emission in the southern region consists of two separate sources: the extended SouthLobe and the point-like SrcX. Among them, SouthLobe has a flat spectrum with an index of $\sim$ 2.0. Considering that there is a good spatial correspondence between the southern $\gamma$-ray emission and molecular cloud distribution shown in Fig. \ref{fig:4}, the SouthLobe could originate 
from interaction of accelerated CRs with MCs. A number of arguments suggest that SrcX can be interpreted as 
molecular clouds illuminated by escaping CRs with their projection located within the shell of the SNR \citep{2023ApJ...953..100L}. First, for SrcX,
there is no detected X-ray emission and no pulse signal\citep{2020A&A...643A..28D}. Moreover, it is in spatial coincidence with molecular clouds and the high density environment calculated in Sect. \ref{sec:co}. Finally, the fact that its GeV spectrum index is much softer than others region from the SNR itself, and that its flux is so stable (as shown in Sect. \ref{sec:flux}) further support this scenario. The parameters adopted for this scenario are summarized in Table \ref{table:3}.

To fit the $\gamma$-ray spectra of SouthLobe and SrcX. We assume an instantaneous injection of protons into an uniform emission zone at T = 13 kyr ago.
The injected proton spectrum is a broken power-law:
\begin{equation}
Q(E) = {Q_0} \frac{(E/E_{p,\rm br})^{-\gamma_{1}}}{1+(E/E_{p,\rm br})^{\gamma_{2}-\gamma_{1}}}\,. 
\label{eq:p_spectra}
\end{equation}
Here $\gamma_{1}$ and $\gamma_{2}$ are the spectral indices below and above the break energy $E_{p,\rm br}$, respectively. A part of the injected protons will be trapped inside while others can escape via diffusion. Such a broken power law spectrum is often seen in $\gamma$-ray SNRs \citep{2017ApJ...834..153Z,2019ApJ...874...50Z,2021ApJ...910...78Z,2022ApJ...928...89H,2023ApJ...953..100L}. Given the flat $\gamma$-ray spectrum in GeV energy band for SouthLobe, we have $\gamma_{1}$ = 2.0 and $\gamma_{2}$ = $\gamma_{1}$ + 1, the break energy of protons $E_{p, \rm br}$ is fixed at 10 TeV. 

\begin{figure*}
    \centering
    \includegraphics[trim={0 0.cm 0 0}, clip, width=0.35\textwidth]{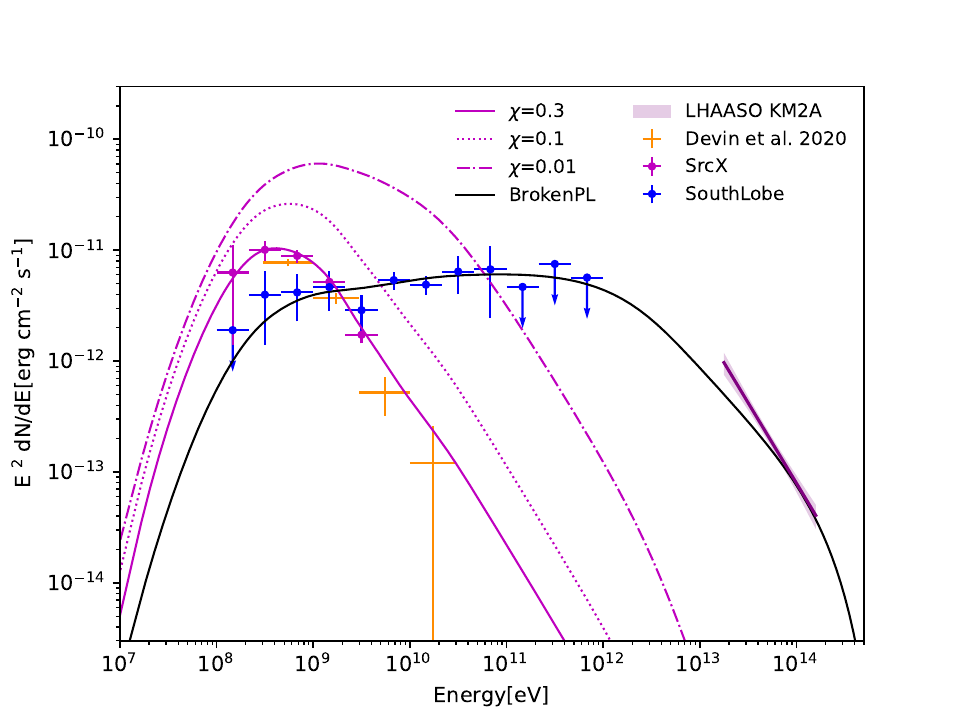}
    \includegraphics[trim={0 0.cm 0 0}, clip, width=0.35\textwidth]{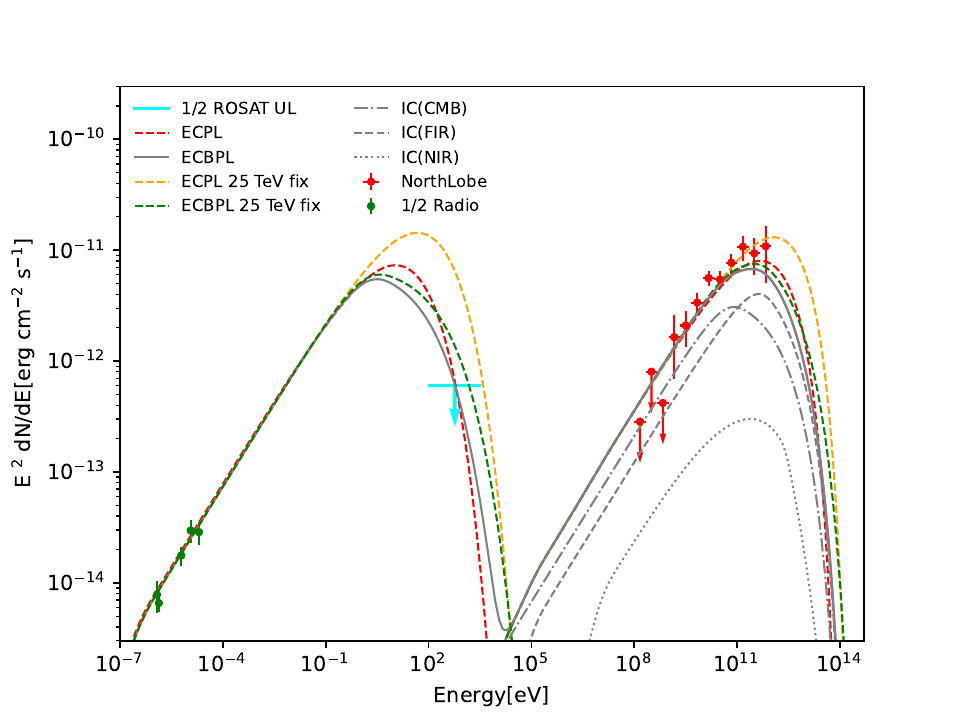} \\
    \includegraphics[trim={0 0.cm 0 0}, clip, width=0.35\textwidth]{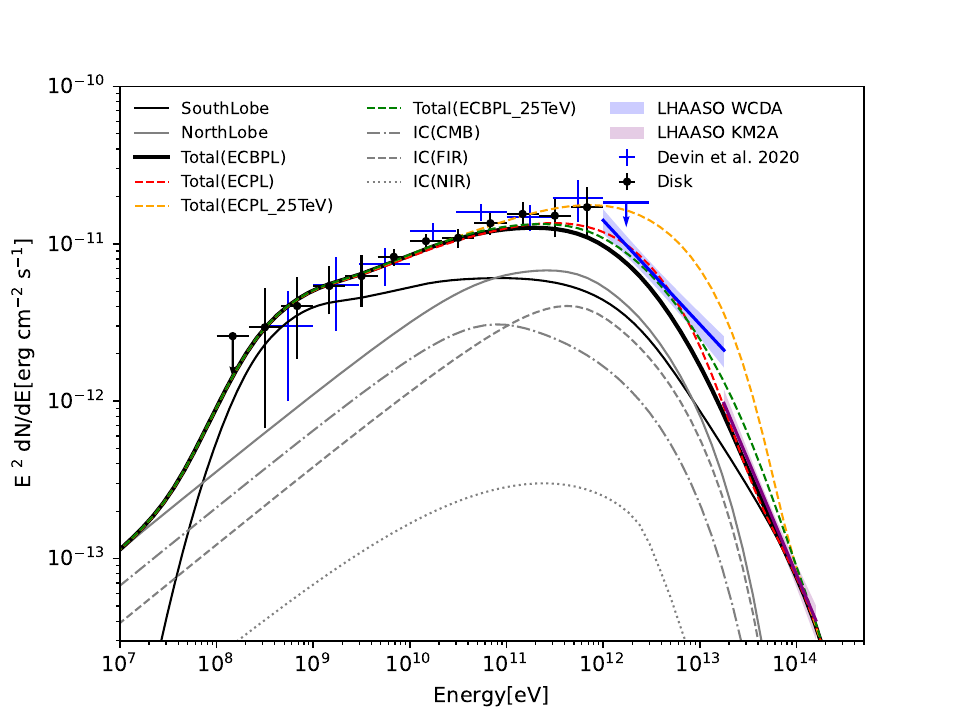}
    \includegraphics[trim={0 0.cm 0 0}, clip, width=0.35\textwidth]{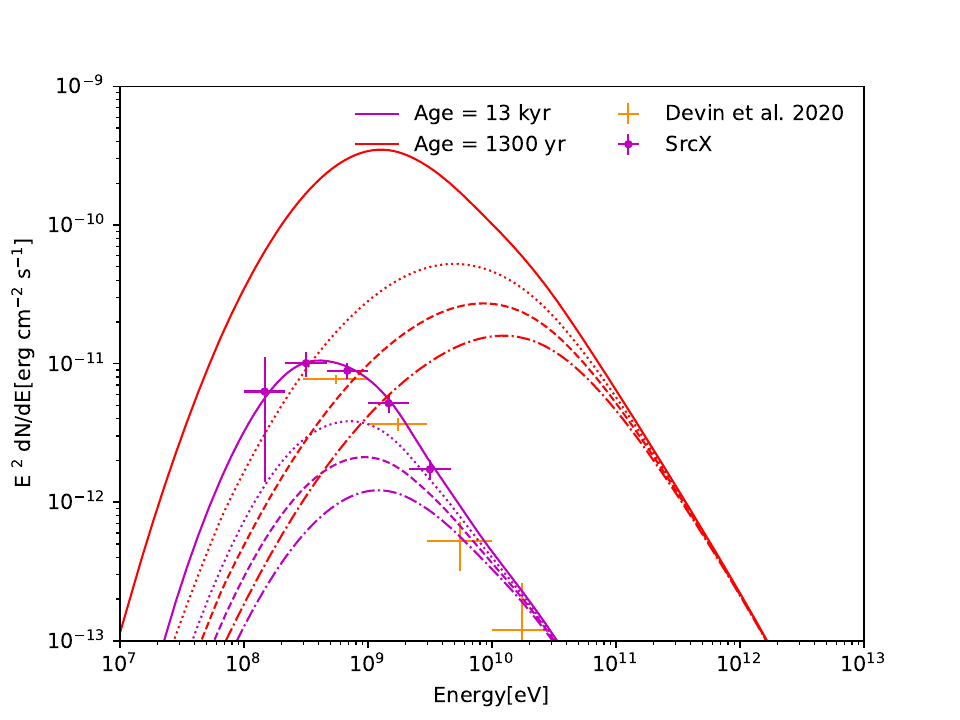}
    \caption{Modeling of the multi-wavelength spectra of the SrcX/SouthLobe (top left), NorthLobe (top right), Disk (bottom left). The blue and purple butterfly shadow regions come from recent LHAASO results\citep{2023arXiv230517030C}. The orange data points come from 4FGL J0426.5+5434 as measured by \citet{2020A&A...643A..28D}. Top left: Solid, dashed and dash-dotted magenta lines show the spectra produced by escaping ions with different values of $\chi$ indicated, and the black solid line shows the hadronic model for trapped protons with a broken power-law distribution. Top right: Leptonic models, where the power-law (ECPL) and broken power-law (ECBPL) are presented by red dashed and grey solid lines, and the orange/green dashed lines show a special ECPL/ECBPL case where the cutoff energy is fixed to 25 TeV. The radio fluxes and the X-ray upper-limit come from \citet{2021ApJ...910...78Z} and \citet{2020A&A...643A..28D}, respectively. Bottom left: Hybrid model for the total $\gamma$-ray flux shown as black solid/red dashed/orange dashed lines added by different leptonic spectral types. Blue data points are derived from the best-fit results measured by \citet{2020A&A...643A..28D} for comparison. Bottom right: Dependence of the SED on the SNR age and r$_{\rm s}$.
    The solid, dotted, dashed, dash-dotted lines correspond to r$_{\rm s}$ = 21.4/35/45/55 pc, respectively.}
    \label{fig:5}
\end{figure*}

Assuming that the shock-cloud interaction site is far away from the shock front and CRs is released at the time of the SN explosion, the distribution of the escaped protons follows \citep{2012MNRAS.419..624T, 2020ApJ...897L..34L,2023ApJ...945...21L}:
\begin{equation}
N_p(E, r_{\rm s}, T)=\frac{Q(E)}{[4 \pi D(E) T]^\frac{3}{2}}  \exp\left[\frac{-r_{\rm s}^2}{4 D(E) T}\right]\label{equation:3}
\end{equation}
where the diffusion coefficient is assumed to be uniform and taken to be $D(E)=\chi D_0(E/E_0)^\delta$ for $E>E_0$, where $D_0=1\times 10^{28}$ cm$^2$ s$^{-1}$ at $E_0=10$ GeV and $\delta$ = 1. 
With a distance between the Earth and the SNR G150.3+4.5 $d = 0.8$ kpc, the $\gamma$-ray SNR has a radius R = $\theta \times$d = 21.4 pc, the same as in Sect. \ref{sec:co}. Considering the physical distance between SrcX and SNR central can not be constrained by observations for the projection effect, we assume r$_{\rm s}$ as the distance between the injection site (SNR centre) and the emitting molecular clouds in the subsequent model building part to fit the SED of SrcX. 
For an injected source spectrum given by $Q(E) \propto E^{-\Gamma}$ and $D(E) \propto E^\delta$, equation \ref{equation:3} shows that $N_p(E)$ will follow $N_p(E) \propto E^{-\left(\Gamma+\frac{3}{2}\delta\right)}$ at high energies, while displaying a lower energy spectral cutoff at E$_b$ where $\sqrt{4D(E_b)T} \simeq r_{\rm s}$. The bottom right panel of Fig. \ref{fig:5} shows the dependence of the SED on the SNR age and $r_{\rm s}$. The $\sim21$ pc (magnetic solid line) scenario is perfectly compatible with the SED trend, while under the larger $r_{\rm s}$ assumption, the peak values of the fit lines are difficult to match with the observed data points (even with a higher injection energy). Thus we propose that there is such a constraint on the approximate value of $r_{\rm s}$.

The SED trend is perfectly compatible with the 21 pc (magnetic solid line) scenario, but the peak values of the fit lines under the bigger $r_{\rm s}$ assumption are difficult to match the observed data (even with a greater inject energy). For this reason, we propose that there is a constraint regarding the approximate value of Rs.

The total energy of injected protons is assumed to be W$_{\rm inj}$ = $\eta$ E$_{\rm SN}$, the ratio $\eta$ of the kinetic energy converted into accelerated protons with a standard value of 0.1, the E$_{\rm SN}$ is the kinetic energy of the SNR with a typical value of 10$^{51}$erg. The correction factor $\chi$ of the diffusion coefficient is set to be free parameter. 
The corresponding $\gamma$-ray fluxes are calculated using the naima package \citep{zabalza2015naima} with:
\begin{equation}
    \frac{dN_\gamma}{dE_\gamma}=\frac{n_i v_i c }{4\pi d^2} \int \frac{d\sigma_{pp}}{dE_\gamma}\left(E_\gamma,E\right)N_{\rm p}(E,r)dE 
\end{equation}
\\
where the differential proton-proton inelastic cross section for $\gamma$-ray production, $d\sigma_{pp}/dE_{\gamma}$, is adopted from \citet{PhysRevD.90.123014}. $n_i$/$v_i$ represents for the total gas density/volume in different emission zones ($i$ = SouthLobe, X) and are the same as those given in Sect. \ref{sec:co}.
For the southern region, the resulting $\gamma$-ray flux with the parameter $\chi=0.3$ could perfectly explain the observational data for T = 13 kyr as indicated by the magenta line in the top left panel of Fig. \ref{fig:5}. 
And the total energy of escaped protons above 1 GeV in the X emission zone is calculated to be W$_{\rm escaped,X} = 2.21\times 10^{47}
(n_{\rm X}/89 {\rm cm}^{-3})^{-1}$  erg, while the total energy of trapped protons above 1 GeV in the SouthLobe emission zone is calculated to be W$_{\rm trapped} = 2.35\times 10^{47} (n_{\rm S}/66 {\rm cm}^{-3})^{-1}$ erg. We note that the diffusion coefficient is one order of magnitude lower than the standard Galactic value, and a smaller diffusion coefficient (e.g., $\chi=0.1$) would leads to a lower value of W$_{\rm escaped,X}$ to explain the observed flux. The soft GeV spectrum however requires a value of $\delta$ as high as 1, 
corresponding to Bohm-like diffusion, which is different from typical turbulence models such as Kolmogorov and Kraichnan found in others SNRs. The presence of Bohm diffusion around this SNR could be due to CR-driven instabilities from CRs leaving that source, and modifying the magnetic field in the ISM around the SNR \citep{2004MNRAS.353..550B}. It is known from numerical simulations \citep{2013MNRAS.430.2873R} that the magnetic field in the upstream of SNR shocks leads to Bohm diffusion for CRs. It is therefore plausible that CRs could undergo Bohm-like diffusion around their SNR. A similar diffusion coefficient was also found in HESS J1912+101 \citep{2023ApJ...953..100L}. 

\begin{table*}[h]  
    \caption{Fiducial parameters for hadronic model} \label{tab:modelpara}
    \centering
    \begin{tabular}{lccc}
    \hline
    SNR parameters &  Symbol &  &\\\hline
    SN explosion energy & $E_{\rm SN}$ & \multicolumn{2}{c}{$1.0\times10^{51}~{\rm erg}$}    \\ 
    Convert efficiency & $\eta$ & \multicolumn{2}{c}{$0.1$ } \\
    Age of the SNR & $T$ & \multicolumn{2}{c}{$13.0~{\rm kyr}$}    \\
    Distance to the SNR & $d$ & \multicolumn{2}{c}{$0.8~{\rm kpc}$}    \\ 
    SNR radius & $R$ & \multicolumn{2}{c}{$ 21.4~{\rm pc}$}\\
    Energy break & $E_{p,\rm br}$ & \multicolumn{2}{c}{$10~{\rm TeV}$}     \\
    Particle index trapped in the SNR & $Broken Power-Law$ & \multicolumn{2}{c}{2.0/3.0}    \\
    Particle index escaped from SNR & $N_p(E, r_{\rm s}, T)$ & \multicolumn{2}{c}{2.0 + 1.5 $\delta$}    \\
    \hline  Diffusion parameters &  Symbol &  &  \\\hline
    Diffusion coefficient at $E=10~{\rm GeV}$ & $D_0$ & \multicolumn{2}{c}{$1.0 \times 10^{28}~{\rm cm}^2~{\rm s}^{-1}$}    \\ 
    Index of dependence on $E$ of diffusion & $\delta$ & \multicolumn{2}{c}{$1.0$}    \\ 
    Factor of dependence on $E$ of diffusion & $\chi$ & \multicolumn{2}{c}{$0.3/0.1/0.01$}    \\ 
    \hline  Molecular cloud (MC) parameters &  Symbol & SrcX(0.1$^{\circ}\!$ sphere) & SouthLobe  \\\hline
    Distance to the MC from SNR & r$_{\rm s}$  & $21.4/35/45/55~{\rm pc}$&$-$\\
    Average hydrogen number density& $n_{\rm H}$ & $~89{\rm cm}^{-3}$ & $66{\rm cm}^{-3}$  \\
    Mass of the  MC& $M_{\rm cl}$ & $2.5\times10^{2}~M_\odot$ & $3.1\times10^{4}~M_\odot$ \\
    \hline
    \label{table:3}
    \end{tabular}
\end{table*}

\subsection{Leptonic emission dominated in the northern region }
On the inner side of the shock front, the $\gamma$-ray emission is assumed to be mainly from the shock-accelerated electrons, which are also producing the synchrotron emission in the spectral range from radio to X-rays, mainly owing to the consistency with the hard GeV $\gamma$-ray spectrum similar to typical PWN such as HESS J1825-137\citep{2020A&A...640A..76P}, HESS J1640-465 \citep{2021ApJ...912..158M}, etc. 
The radio spectral index $-0.69 \leq \alpha \leq -0.40$ \citep{2014A&A...567A..59G} in the SNR G150 region is much softer than that of a typical PWN ($-0.3 \lesssim \alpha \lesssim 0$, \cite{2006ARA&A..44...17G}) and is more compatible with a SNR scenario ($\alpha \lesssim -0.40$, \cite{2017yCat.7278....0G}). Moreover, there is no pulsation \citep{2013MNRAS.429.1633B}, nor X-ray emission detected \citep{2020A&A...643A..28D}, and the radio shell structure shown in Fig. \ref{fig:1} is very symmetric. Therefore, we suggest that the $\gamma$-ray emission in the NorthLobe also comes from SNR G150 and is dominated by the inverse-Compton scattering process. In this case, we performed a multi-wavelength modeling using the naima package \citep{zabalza2015naima} with a simple one-zone leptonic model for the NorthLobe with a Markov Chain Monte Carlo algorithm to give the numerical constraint. The best-fit parameters are summarized in Table \ref{table:SrcN}. On the other hand, we note the potential irregular and elongated structure in the northern region, indicating the possibility of (filamentary) diffusion caused by local magnetic turbulence \citep{2012PhRvL.108z1101G}, thus the high-energy electrons escape scenario can not be ruled out. Combined with the age of SNR G150.3+4.5 in the ranges 10 - 100 kyr, there is accord with potential pulsar halo evolution stage 2 depicted in \citet{2020A&A...636A.113G}, which means that under this assumption, the potential pulsar halo/PWN hasn't yet fully escaped from the SNR region. Thus the $\gamma$-ray emission might come from a SNR plus PWN contribution \citep{Zeng:2023uvu}. Such a scenario cannot be totally ruled out, and more high-resolution observations are needed to reveal the nature in this region.

Some evidence was shown that the radio continuum spectral indices 
for the northern and southern parts of G150.3+4.5 might be different 
\citep{2014A&A...567A..59G}. We therefore treat the two parts 
separately. According to the procedure introduced in 
\citet{Gao11}, the large-scale Galactic diffuse emission in the 
observation image is filtered out by using the technique 
of ``background filtering'' \citep{Sofue79}. For a further adjustment, 
a twisted hyper-plane is constructed based on the pixel values 
in the four image corners without any obvious structures and subtracted 
to remove the local-background Galactic diffuse emission, which 
ensures that the radio continuum emission seen in the target area 
all comes from G150.3+4.5. Finally, we estimate the flux 
densities for the two parts which are separated by the white 
line as marked in Fig.~\ref{fig:1}. Firstly, the above procedure is applied
to the Effelsberg 21\ cm image \citep{Reich97} as shown in the Fig.~1 of \citet{2014A&A...567A..59G}. 
The results are $S_{\rm 21cm} = 5.1$~Jy 
and $S_{\rm 21cm} = 6.3$~Jy for the northern and southern parts of G150.3+4.5, respectively. 
The flux density at 21\ cm for the northern part ($\sim$5.1~Jy) could be regarded as a lower 
limit, since some faint emission seen at 6\ cm in this part 
is not detected at 21\ cm. Then the same routine is further applied to Urumqi 6\ cm data, 
and we obtain $S_{\rm 6cm} = 3.3$~Jy, 
and $S_{\rm 6cm} = 3.5$~Jy accordingly. Considering the uncertainty, i.e. 10\% in the measurements \citep{2014A&A...567A..59G}, 
the radio flux densities given by Urumqi 6\ cm in both parts are almost equal. Thus we adopted 0.5 times radio flux for NorthLobe model building 
calculation shown in Fig. \ref{fig:5} top right panel.

In the leptonic model, the inverse Compton scattering (ICS) or bremsstrahlung processes of relativistic electrons is considered, especially for the ICS process, the radiation field includes the CMB and IR component from interstellar dust and gas with T = 30 K and u = 1 eV cm$^{-3}$ \citep{2006ApJ...648L..29P,2008ApJ...682..400P}. The distance is also adopted as 0.8 kpc. The spectra of the electrons was assumed to be power-law with an exponential cutoff (ECPL) in the form of:
\begin{equation}
    \frac{dN_{\rm e}}{dE} \propto \left(\frac{E}{E_0}\right)^{-\alpha_{e1}} exp\left(-\frac{E}{E_{\rm e, cut}}\right) 
\end{equation}
and broken power-law (ECBPL) with an exponential cutoff, which follows\citep{2019ApJ...874...50Z}:

\begin{equation}
\resizebox{0.92\hsize}{!}{$
\frac{dN_{\rm e}}{dE} \propto exp\left(-\frac{E}{E_{\rm e,cut}}\right) 
\begin{cases}
\left(\frac{E}{E_0}\right)^{-\alpha_{e1}} \qquad \qquad \qquad \qquad;  E < E_{\rm e,break}  \\
\left(\frac{E_{\rm e,break}}{E_0}\right)^{\alpha_{e2}-\alpha_{e1}} \left(\frac{E}{E_0}\right)^{-\alpha_{e2}} \,\,\,\,; E \geq E_{\rm e,break}
\end{cases}
$}
\end{equation}
where $\alpha_{\rm e}$ and $\rm E_{e, cut}$ are the spectral index and the cutoff energy, $E_{\rm e,break}$ represents the break energy, and $\alpha_{e2}$ = $\alpha_{e1}$ + 1.0. Then we set the synchrotron cooling timescale being equal to the age of SNR G150.3+4.5 to reduce the number of free parameters, at the cutoff energy E$_{\rm e, cut}$ = 1.25 $\times$ 10$^{\rm 7}$ t$_{\rm age; yr}^{-1}$ B$_{\rm \mu G}^{\rm -2}$ TeV, and combine with the radio flux data points and the upper limit in X-ray to constrain the total energy of electrons above 1 GeV and magnetic field strength, which is calculated to be W$_{\rm e}$ = 1.73 $\times$ 10$^{\rm 47}$ erg and B $\simeq$ 7.0 $\mu$G in broken power-law case, the corresponding $\rm E_{e, cut}$ is calculated to be 19.6\,TeV. For the power-law case (red dashed line), we derived similar results that W$_{\rm e}$ = 1.75 $\times$ 10$^{\rm 47}$ erg and B $\simeq$ 7.6 $\mu$G, $\rm E_{e, cut}$ to be 16.5\,TeV. For the PL (BPL) cases with fixed cutoff energy equal to 25\,TeV, W$_{\rm e}$ = 2.02 (1.74) $\times 10^{\rm 47}$\,erg and the magnetic strength is calculated as B $\simeq 6.2\,\mu$G. Based on the above best-fit parameters, we derived a similar magnetic strength value compared with \citet{2020A&A...643A..28D}. However, due to the ambient gas densities being calculated from different surveys, the estimated SNR age varies, and we suggest that using gas detector measurements will provide a more accurate density estimate than X-ray measurements. Since the best-fit results for freshly accelerated electrons and protons is measured in each lobe, the electron-to-proton ratio can be calculated as $K_{ep}$ = 6 $\times$ 10$^{-3}$ at 1 GeV in the whole SNR region. This is consistent with the expectation in the range of 10$^{-5}$ - 10$^{-2}$ \citep{2013MNRAS.434.2748C,2021MNRAS.508.2204C}, and the typical value of $K_{ep}$ = 10$^{-2}$ was derived from the CRs measured at Earth \citep{2016ApJ...821...43Z}. Also a lower limit of 10$^{-3}$ was obtained according to the radio observations of the SNRs in nearby galaxies \citep{2008JCAP...01..018K}.

\begin{table*}
	\centering
	\caption {Parameters for leptonic dominated NorthLobe}
	\begin{tabular}{ccccccc}
		\hline \hline
		Model & $\alpha_e1$ & $\alpha_e2$ & E$_{e,\rm break}$ & E$_{e,\rm cut}$ & W$_e$ & $B$ \\
		      &             &             & (TeV)   & (TeV) & ($10^{47}$ erg)  & ($\mu$G)\\
		ECPL    & 2.0         & $-$         & $-$   & 16.5   & $1.75_{-0.14}^{+0.11}$ & $7.62_{-0.49}^{+0.58}$   \\
		ECBPL   & 2.0         & 3.0   & $3.02_{-1.03}^{+0.95}$  & 19.6  & $1.73_{-0.12}^{+0.15}$  & $7.01_{-0.61}^{+0.45}$  \\
        ECPL(E$_{e,\rm cut}$ fix)    & 2.0         & $-$         & $-$   & 25.0   & $2.02_{-0.23}^{+0.26}$ & 6.2   \\
        ECBPL(E$_{e,\rm cut}$ fix)    & 2.0         & 3.0         &  $3.03_{-0.93}^{+1.17}$   & 25.0   & $1.74_{-0.13}^{+0.16}$ & 6.2   \\
		\hline
		\hline
	\end{tabular}
	\label{table:SrcN}\\
    {{\bf{Notes.}}{The total energy of relativistic particles, $W_e$ is calculated for $E > 1$ GeV.}}
\end{table*}

Then we add the $\gamma$-ray contributions from the SouthLobe and the NorthLobe together to derive the total $\gamma$-ray flux in the SNR G150.3+4.5 region, which is shown in the bottom left panel of Fig. \ref{fig:5}. The obtained total $\gamma$-ray flux can well explain the apparent discontinuity at the spectral boundary between WCDA and KM2A given by LHAASO \citep{2023arXiv230517030C}. Notably, the red dashed line, with a lower cutoff energy value, is perfectly consistent with both KM2A and WCDA measurements. In contrast, the black solid line and the orange/green dashed lines, which have higher cutoff energies, do not fit well with the WCDA/KM2A results. This means that the $\gamma$-ray emission in the lower TeV energy band detected by WCDA jointly comes from the hadronic and leptonic components, and, as the energy increases, the leptonic component quickly disappears. In the higher TeV energy band, only the hadronic component remains in the high density region. This leads to the softer spectrum detected by KM2A, as well as the smaller extended source, which can smoothly connect with the SED from the SouthLobe. This evidence suggest that SrcX is not the GeV counterpart of the KM2A source, even though they have similar positions. Correspondingly, the KM2A source should be associated with SouthLobe. We also noticed that the disappearance of the leptonic components is limited by multi-wavelength data. The leptonic model can match with LHAASO observations only when the cutoff energy is below the KM2A energy band ($\textless$ 25 TeV). Therefore, this conclusion is to some extent constrained by observations. Conducting further theoretical research about the mechanism of electron cooling in this region is worthwhile.

\section{Discussion and conclusions}\label{sec:5}
 
We analysed the MeV-GeV $\gamma$-ray emission in the vicinity of SNR G150.3+4.5 using 14 years of \emph{Fermi}-LAT data and found that the northern half-sphere 
has a spectral break near 146 GeV with a significance of $\rm{TS}\sim$24 (4.9 $\sigma$). The photon indices below and above the energy break are $1.54\pm0.04_{\rm{stat}}\pm0.07_{\rm{syst}}$ and $2.28\pm0.08_{\rm{stat}}\pm0.12_{\rm{syst}}$, respectively. The southern half-sphere could be described well by a single power-law with an index of $1.95\pm0.07_{\rm{stat}}\pm0.09_{\rm{syst}}$ even though emission above 100 GeV is insignificant. Molecular clouds are only found in the southern half-sphere, implying soft GeV spectra in high density regions, a feature that has been observed in RX J1713.7-3946 \citep{2013ApJ...778...59S, 2020ApJ...900L...5T}, SNR Puppis A \citep{2017ApJ...843...90X}, and
recently HESS J1912+101 \citep{2023ApJ...953..100L}.
In light of the discovery of spectral evolution of high energy particles in SNRs \citep{2017ApJ...834..153Z, 2019ApJ...874...50Z}, these results suggest that shocks evolving in a high density environment produce high energy particle distribution similar to those in old SNRs.

The northern half-sphere has a soft radio spectrum \citep{2014A&A...567A..59G} and the multi-wavelength SED can be fitted with the leptonic scenario for the $\gamma$-ray emission, under ECPL/ECBPL spectral assumption, the total energy of electrons above 1 GeV is calculated as 1.75/1.73 $\times$ 10$^{\rm 47}$ erg, respectively, both of them reveal a high energy cutoff around several tens TeV, which is consistent with measurements towards the whole SNR region \citep{Zeng:2023uvu}. The bright \emph{Fermi} source SrcX (4FGL J0426.5+5434) has a soft spectrum and is spatially coincident with a puzzling 
source discovered by LHAASO-KM2A and a bright radio arc with a hard spectrum  in the southern half-sphere \citep{2014A&A...566A..76G, 2014A&A...567A..59G}. It can be attributed to molecular clouds illuminated by CRs injected at the birth of the SNR,
and its SED can be fitted by a model with the typical SNR
kinetic energy E$_{\rm SN}$ = 10$^{51}$erg and energy convert fraction $\eta$= 0.1, while the best-fit diffusion coefficient is about one order of magnitude lower than the standard Galactic value ($\chi$ = 0.3).
We found that the multi-wavelength SED of the southern half-sphere can be fitted with the hadronic scenario for the $\gamma$-ray emission with the emitting ions having a broken power-law spectrum, reminiscence spectra of middle age SNRs. The $\gamma$-ray emission detected by the LHAASO-KM2A is then produced by CRs interacting with ions in the background. Although the soft spectrum of SrcX may be attributed to CRs escaping from the SNR and interacting with a dense MCs, it may also be associated with a soft spectrum of CRs accelerated as the shock of SNR G150.3+4.5 encounters a dense MCs and slows down \citep{2023ApJ...953..100L}. Further observations are needed to reveal of the nature of SrcX. 

Overall, we found that the GeV spectral indices differ significantly between the north and south sides of the SNR G150.3 + 4.5. Combining the spatial correspondence between the distribution of MCs and the $\gamma$-ray emission in southern region, we suggest that the softer spectral index in the high-density southern region originates from hadronic emission processes. The puzzling point-like source SrcX can be explained by an escape model. In the northern region, the harder spectrum is dominated by the leptonic inverse Compton scattering process. Combining multi-wavelength data, we constrained the fitting parameters for each source in this region. The fitting results show that the hybrid model with both leptonic and hadronic contributions to the $\gamma$-rays is more reasonable than a purely leptonic model. We suggest that the hybrid $\gamma$-ray emission 
is mainly due to evolution of the SNR in an inhomogeneous environment. 
Shocks sweeping through the low-density northern region lead to a harder spectral index dominated by the leptonic emission, while the high-density southern region has a softer spectral index dominated by the hadronic component. In higher energy bands probed by LHAASO-KM2A, only high-density regions bombarded by high energy CRs can be observed.
\begin{acknowledgements}
       We thank Shaoqiang Xi for the help on the molecular cloud data analysis approach. We thank Xuyang Gao for the help on the radio continuum data based on the Urumqi and Effelsberg observations. We also would like to thank Houdun Zeng, Yang Su, Xi Liu and P.P.Delia for invaluable discussions. This work is supported by the National Natural Science Foundation of China under the grants No. 12393853, 12375103, U1931204, 12103040, 12147208, and 12350610239, the Natural Science Foundation for Young Scholars of Sichuan Province, China (No. 2022NSFSC1808), and the , Fundamental Research Funds for the Central Universities (No. 2682022ZTPY013).
\end{acknowledgements}

\bibliographystyle{aa}
\bibliography{ref}

\begin{thebibliography}{82}
\expandafter\ifx\csname natexlab\endcsname\relax\def\natexlab#1{#1}\fi

\bibitem[{{Abdo} {et~al.}(2010){Abdo}, {Ackermann}, {Ajello}, {Allafort},
  {Asano}, {Baldini}, {Ballet}, {Barbiellini}, {Baring}, {Bastieri}, {Bechtol},
  {Bellazzini}, {Berenji}, {Blandford}, {Bloom}, {Bonamente}, {Borgland},
  {Bregeon}, {Brez}, {Brigida}, {Bruel}, {Buson}, {Caliandro}, {Cameron},
  {Camilo}, {Caraveo}, {Carrigan}, {Casandjian}, {Cecchi}, {{\c{C}}elik},
  {Chekhtman}, {Cheung}, {Chiang}, {Ciprini}, {Claus}, {Cohen-Tanugi},
  {Conrad}, {den Hartog}, {Dermer}, {de Luca}, {de Palma}, {Dormody}, {Silva},
  {Drell}, {Dubois}, {Dumora}, {Farnier}, {Favuzzi}, {Fegan}, {Ferrara},
  {Focke}, {Frailis}, {Fukazawa}, {Funk}, {Fusco}, {Gargano}, {Gehrels},
  {Germani}, {Giglietto}, {Giordano}, {Glanzman}, {Godfrey}, {Gotthelf},
  {Grenier}, {Grondin}, {Grove}, {Guillemot}, {Guiriec}, {Hanabata}, {Harding},
  {Hays}, {Hobbs}, {Horan}, {Hughes}, {J{\'o}hannesson}, {Johnson}, {Johnson},
  {Johnson}, {Johnston}, {Kamae}, {Kanai}, {Kanbach}, {Katagiri}, {Kataoka},
  {Kawai}, {Keith}, {Kerr}, {Kn{\"o}dlseder}, {Kuss}, {Lande}, {Latronico},
  {Lemoine-Goumard}, {Llena Garde}, {Longo}, {Loparco}, {Lott}, {Lovellette},
  {Lubrano}, {Makeev}, {Manchester}, {Marelli}, {Mazziotta}, {McEnery},
  {Michelson}, {Mitthumsiri}, {Mizuno}, {Moiseev}, {Monte}, {Monzani},
  {Morselli}, {Moskalenko}, {Murgia}, {Nakamori}, {Nolan}, {Norris}, {Nuss},
  {Ohno}, {Ohsugi}, {Omodei}, {Orlando}, {Ormes}, {Paneque}, {Panetta},
  {Parent}, {Pelassa}, {Pepe}, {Pesce-Rollins}, {Piron}, {Porter}, {Rain{\`o}},
  {Rando}, {Razzano}, {Rea}, {Reimer}, {Reimer}, {Reposeur}, {Rodriguez},
  {Romani}, {Roth}, {Ryde}, {Sadrozinski}, {Sander}, {Saz Parkinson},
  {Sgr{\`o}}, {Siskind}, {Smith}, {Smith}, {Spandre}, {Spinelli}, {Starck},
  {Strickman}, {Suson}, {Takahashi}, {Takahashi}, {Tanaka}, {Thayer}, {Thayer},
  {Thompson}, {Thorsett}, {Tibaldo}, {Torres}, {Tosti}, {Tramacere},
  {Uchiyama}, {Usher}, {Vasileiou}, {Venter}, {Vilchez}, {Vitale}, {Waite},
  {Wang}, {Weltevrede}, {Winer}, {Wood}, {Yang}, {Ylinen}, {Ziegler}, {Fermi
  LAT Collaboration}, \& {Pulsar Timing Consortium}}]{2010ApJ...714..927A}
{Abdo}, A.~A., {Ackermann}, M., {Ajello}, M., {et~al.} 2010, \apj, 714, 927

\bibitem[{{Abdollahi} {et~al.}(2020{\natexlab{a}}){Abdollahi}, {Acero},
  {Ackermann}, {Ajello}, {Atwood}, {Axelsson}, {Baldini}, {Ballet},
  {Barbiellini}, {Bastieri}, {Becerra Gonzalez}, {Bellazzini}, {Berretta},
  {Bissaldi}, {Blandford}, {Bloom}, {Bonino}, {Bottacini}, {Brandt}, {Bregeon},
  {Bruel}, {Buehler}, {Burnett}, {Buson}, {Cameron}, {Caputo}, {Caraveo},
  {Casandjian}, {Castro}, {Cavazzuti}, {Charles}, {Chaty}, {Chen}, {Cheung},
  {Chiaro}, {Ciprini}, {Cohen-Tanugi}, {Cominsky}, {Coronado-Bl{\'a}zquez},
  {Costantin}, {Cuoco}, {Cutini}, {D'Ammando}, {DeKlotz}, {de la Torre Luque},
  {de Palma}, {Desai}, {Digel}, {Di Lalla}, {Di Mauro}, {Di Venere},
  {Dom{\'\i}nguez}, {Dumora}, {Fana Dirirsa}, {Fegan}, {Ferrara},
  {Franckowiak}, {Fukazawa}, {Funk}, {Fusco}, {Gargano}, {Gasparrini},
  {Giglietto}, {Giommi}, {Giordano}, {Giroletti}, {Glanzman}, {Green},
  {Grenier}, {Griffin}, {Grondin}, {Grove}, {Guiriec}, {Harding}, {Hayashi},
  {Hays}, {Hewitt}, {Horan}, {J{\'o}hannesson}, {Johnson}, {Kamae}, {Kerr},
  {Kocevski}, {Kovac'evic'}, {Kuss}, {Landriu}, {Larsson}, {Latronico},
  {Lemoine-Goumard}, {Li}, {Liodakis}, {Longo}, {Loparco}, {Lott},
  {Lovellette}, {Lubrano}, {Madejski}, {Maldera}, {Malyshev}, {Manfreda},
  {Marchesini}, {Marcotulli}, {Mart{\'\i}-Devesa}, {Martin}, {Massaro},
  {Mazziotta}, {McEnery}, {Mereu}, {Meyer}, {Michelson}, {Mirabal}, {Mizuno},
  {Monzani}, {Morselli}, {Moskalenko}, {Negro}, {Nuss}, {Ojha}, {Omodei},
  {Orienti}, {Orlando}, {Ormes}, {Palatiello}, {Paliya}, {Paneque}, {Pei},
  {Pe{\~n}a-Herazo}, {Perkins}, {Persic}, {Pesce-Rollins}, {Petrosian},
  {Petrov}, {Piron}, {Poon}, {Porter}, {Principe}, {Rain{\`o}}, {Rando},
  {Razzano}, {Razzaque}, {Reimer}, {Reimer}, {Remy}, {Reposeur}, {Romani}, {Saz
  Parkinson}, {Schinzel}, {Serini}, {Sgr{\`o}}, {Siskind}, {Smith}, {Spandre},
  {Spinelli}, {Strong}, {Suson}, {Tajima}, {Takahashi}, {Tak}, {Thayer},
  {Thompson}, {Tibaldo}, {Torres}, {Torresi}, {Valverde}, {Van Klaveren}, {van
  Zyl}, {Wood}, {Yassine}, \& {Zaharijas}}]{2020ApJS..247...33A}
{Abdollahi}, S., {Acero}, F., {Ackermann}, M., {et~al.} 2020{\natexlab{a}},
  \apjs, 247, 33

\bibitem[{{Abdollahi} {et~al.}(2020{\natexlab{b}}){Abdollahi}, {Acero},
  {Ackermann}, {Ajello}, {Atwood}, {Axelsson}, {Baldini}, {Ballet},
  {Barbiellini}, {Bastieri}, {Becerra Gonzalez}, {Bellazzini}, {Berretta},
  {Bissaldi}, {Blandford}, {Bloom}, {Bonino}, {Bottacini}, {Brandt}, {Bregeon},
  {Bruel}, {Buehler}, {Burnett}, {Buson}, {Cameron}, {Caputo}, {Caraveo},
  {Casandjian}, {Castro}, {Cavazzuti}, {Charles}, {Chaty}, {Chen}, {Cheung},
  {Chiaro}, {Ciprini}, {Cohen-Tanugi}, {Cominsky}, {Coronado-Bl{\'a}zquez},
  {Costantin}, {Cuoco}, {Cutini}, {D'Ammando}, {DeKlotz}, {de la Torre Luque},
  {de Palma}, {Desai}, {Digel}, {Di Lalla}, {Di Mauro}, {Di Venere},
  {Dom{\'\i}nguez}, {Dumora}, {Fana Dirirsa}, {Fegan}, {Ferrara},
  {Franckowiak}, {Fukazawa}, {Funk}, {Fusco}, {Gargano}, {Gasparrini},
  {Giglietto}, {Giommi}, {Giordano}, {Giroletti}, {Glanzman}, {Green},
  {Grenier}, {Griffin}, {Grondin}, {Grove}, {Guiriec}, {Harding}, {Hayashi},
  {Hays}, {Hewitt}, {Horan}, {J{\'o}hannesson}, {Johnson}, {Kamae}, {Kerr},
  {Kocevski}, {Kovac'evic'}, {Kuss}, {Landriu}, {Larsson}, {Latronico},
  {Lemoine-Goumard}, {Li}, {Liodakis}, {Longo}, {Loparco}, {Lott},
  {Lovellette}, {Lubrano}, {Madejski}, {Maldera}, {Malyshev}, {Manfreda},
  {Marchesini}, {Marcotulli}, {Mart{\'\i}-Devesa}, {Martin}, {Massaro},
  {Mazziotta}, {McEnery}, {Mereu}, {Meyer}, {Michelson}, {Mirabal}, {Mizuno},
  {Monzani}, {Morselli}, {Moskalenko}, {Negro}, {Nuss}, {Ojha}, {Omodei},
  {Orienti}, {Orlando}, {Ormes}, {Palatiello}, {Paliya}, {Paneque}, {Pei},
  {Pe{\~n}a-Herazo}, {Perkins}, {Persic}, {Pesce-Rollins}, {Petrosian},
  {Petrov}, {Piron}, {Poon}, {Porter}, {Principe}, {Rain{\`o}}, {Rando},
  {Razzano}, {Razzaque}, {Reimer}, {Reimer}, {Remy}, {Reposeur}, {Romani}, {Saz
  Parkinson}, {Schinzel}, {Serini}, {Sgr{\`o}}, {Siskind}, {Smith}, {Spandre},
  {Spinelli}, {Strong}, {Suson}, {Tajima}, {Takahashi}, {Tak}, {Thayer},
  {Thompson}, {Tibaldo}, {Torres}, {Torresi}, {Valverde}, {Van Klaveren}, {van
  Zyl}, {Wood}, {Yassine}, \& {Zaharijas}}]{abdollahi2020a}
{Abdollahi}, S., {Acero}, F., {Ackermann}, M., {et~al.} 2020{\natexlab{b}},
  \apjs, 247, 33

\bibitem[{{Abdollahi} {et~al.}(2022){Abdollahi}, {Acero}, {Baldini}, {Ballet},
  {Bastieri}, {Bellazzini}, {Berenji}, {Berretta}, {Bissaldi}, {Blandford},
  {Bloom}, {Bonino}, {Brill}, {Britto}, {Bruel}, {Burnett}, {Buson}, {Cameron},
  {Caputo}, {Caraveo}, {Castro}, {Chaty}, {Cheung}, {Chiaro}, {Cibrario},
  {Ciprini}, {Coronado-Bl{\'a}zquez}, {Crnogorcevic}, {Cutini}, {D'Ammando},
  {De Gaetano}, {Digel}, {Di Lalla}, {Dirirsa}, {Di Venere}, {Dom{\'\i}nguez},
  {Fallah Ramazani}, {Fegan}, {Ferrara}, {Fiori}, {Fleischhack}, {Franckowiak},
  {Fukazawa}, {Funk}, {Fusco}, {Galanti}, {Gammaldi}, {Gargano}, {Garrappa},
  {Gasparrini}, {Giacchino}, {Giglietto}, {Giordano}, {Giroletti}, {Glanzman},
  {Green}, {Grenier}, {Grondin}, {Guillemot}, {Guiriec}, {Gustafsson},
  {Harding}, {Hays}, {Hewitt}, {Horan}, {Hou}, {J{\'o}hannesson}, {Karwin},
  {Kayanoki}, {Kerr}, {Kuss}, {Landriu}, {Larsson}, {Latronico},
  {Lemoine-Goumard}, {Li}, {Liodakis}, {Longo}, {Loparco}, {Lott}, {Lubrano},
  {Maldera}, {Malyshev}, {Manfreda}, {Mart{\'\i}-Devesa}, {Mazziotta}, {Mereu},
  {Meyer}, {Michelson}, {Mirabal}, {Mitthumsiri}, {Mizuno}, {Moiseev},
  {Monzani}, {Morselli}, {Moskalenko}, {Negro}, {Nuss}, {Omodei}, {Orienti},
  {Orlando}, {Paneque}, {Pei}, {Perkins}, {Persic}, {Pesce-Rollins},
  {Petrosian}, {Pillera}, {Poon}, {Porter}, {Principe}, {Rain{\`o}}, {Rando},
  {Rani}, {Razzano}, {Razzaque}, {Reimer}, {Reimer}, {Reposeur},
  {S{\'a}nchez-Conde}, {Saz Parkinson}, {Scotton}, {Serini}, {Sgr{\`o}},
  {Siskind}, {Smith}, {Spandre}, {Spinelli}, {Sueoka}, {Suson}, {Tajima},
  {Tak}, {Thayer}, {Thompson}, {Torres}, {Troja}, {Valverde}, {Wood}, \&
  {Zaharijas}}]{2022ApJS..260...53A}
{Abdollahi}, S., {Acero}, F., {Baldini}, L., {et~al.} 2022, \apjs, 260, 53

\bibitem[{{Abeysekara} {et~al.}(2023){Abeysekara}, {Albert}, {Alfaro},
  {Alvarez}, {{\'A}lvarez}, {Araya}, {Arteaga-Vel{\'a}zquez}, {Arunbabu},
  {Rojas}, {Solares}, {Babu}, {Barber}, {Becerril}, {Belmont-Moreno}, {BenZvi},
  {Blanco}, {Braun}, {Brisbois}, {Caballero-Mora}, {Mart{\'\i}nez},
  {Capistr{\'a}n}, {Carrami{\~n}ana}, {Casanova}, {Castillo}, {Chaparro-Amaro},
  {Cotti}, {Cotzomi}, {de Le{\'o}n}, {de la Fuente}, {de Le{\'o}n}, {De Young},
  {Hernandez}, {Dingus}, {DuVernois}, {Durocher}, {D{\'\i}az-V{\'e}lez},
  {Ellsworth}, {Engel}, {Espinoza}, {Fan}, {Fang}, {Fick}, {Fleischhack},
  {Flores}, {Fraija}, {Garc{\'\i}a-Gonz{\'a}lez}, {Garcia-Torales}, {Garfias},
  {Giacinti}, {Goksu}, {Gonz{\'a}lez}, {Gonz{\'a}lez-Mu{\~n}oz}, {Goodman},
  {Harding}, {Hernandez}, {Hernandez}, {Hinton}, {Hona}, {Huang},
  {Hueyotl-Zahuantitla}, {Hui}, {Humensky}, {H{\"u}ntemeyer}, {Iriarte},
  {Imran}, {Jardin-Blicq}, {Joshi}, {Kaufmann}, {Kieda}, {Kunde}, {Lara},
  {Lauer}, {Lee}, {Lennarz}, {Vargas}, {Linnemann}, {Longinotti}, {Luis-Raya},
  {Lundeen}, {Malone}, {Marandon}, {Marinelli}, {Martinez},
  {Mart{\'\i}nez-Castellanos}, {Mart{\'\i}nez-Castro}, {Mart{\'\i}nez-Huerta},
  {Matthews}, {Miranda-Romagnoli}, {Montaruli}, {Morales-Soto}, {Moreno},
  {Mostaf{\'a}}, {Nayerhoda}, {Nellen}, {Newbold}, {Nisa}, {Noriega-Papaqui},
  {Oceguera-Becerra}, {Olivera-Nieto}, {Omodei}, {Peisker}, {Araujo},
  {P{\'e}rez-P{\'e}rez}, {Ponce}, {Pretz}, {Rho}, {Rosa-Gonz{\'a}lez},
  {Ruiz-Velasco}, {Salazar}, {Salazar-Gallegos}, {Greus}, {Sandoval},
  {Schneider}, {Schoorlemmer}, {Serna-Franco}, {Sinnis}, {Smith}, {Son},
  {Woodle}, {Springer}, {Taboada}, {Tepe}, {Tibolla}, {Tollefson}, {Torres},
  {Torres-Escobedo}, {Turner}, {Ure{\~n}a-Mena}, {Ukwatta}, {Varela},
  {Vargas-Maga{\~n}a}, {Villase{\~n}or}, {Wang}, {Watson}, {Werner},
  {Westerhoff}, {Willox}, {Wisher}, {Wood}, {Yodh}, {Zaborov}, {Zepeda},
  {Zhou}, {historical}, \& {present HAWC Collaboration}}]{2023NIMPA105268253A}
{Abeysekara}, A.~U., {Albert}, A., {Alfaro}, R., {et~al.} 2023, Nuclear
  Instruments and Methods in Physics Research A, 1052, 168253

\bibitem[{{Abeysekara} {et~al.}(2017){Abeysekara}, {Albert}, {Alfaro},
  {Alvarez}, {{\'A}lvarez}, {Arceo}, {Arteaga-Vel{\'a}zquez}, {Ayala Solares},
  {Barber}, {Baughman}, {Bautista-Elivar}, {Becerra Gonzalez}, {Becerril},
  {Belmont-Moreno}, {BenZvi}, {Berley}, {Bernal}, {Braun}, {Brisbois},
  {Caballero-Mora}, {Capistr{\'a}n}, {Carrami{\~n}ana}, {Casanova}, {Castillo},
  {Cotti}, {Cotzomi}, {Couti{\~n}o de Le{\'o}n}, {de la Fuente}, {De Le{\'o}n},
  {Diaz Hernandez}, {Dingus}, {DuVernois}, {D{\'\i}az-V{\'e}lez}, {Ellsworth},
  {Engel}, {Fiorino}, {Fraija}, {Garc{\'\i}a-Gonz{\'a}lez}, {Garfias},
  {Gerhardt}, {Gonz{\'a}lez Mu{\~n}oz}, {Gonz{\'a}lez}, {Goodman},
  {Hampel-Arias}, {Harding}, {Hernandez}, {Hernandez-Almada}, {Hinton}, {Hui},
  {H{\"u}ntemeyer}, {Iriarte}, {Jardin-Blicq}, {Joshi}, {Kaufmann}, {Kieda},
  {Lara}, {Lauer}, {Lee}, {Lennarz}, {Le{\'o}n Vargas}, {Linnemann},
  {Longinotti}, {Raya}, {Luna-Garc{\'\i}a}, {L{\'o}pez-Coto}, {Malone},
  {Marinelli}, {Martinez}, {Martinez-Castellanos}, {Mart{\'\i}nez-Castro},
  {Mart{\'\i}nez-Huerta}, {Matthews}, {Miranda-Romagnoli}, {Moreno},
  {Mostaf{\'a}}, {Nellen}, {Newbold}, {Nisa}, {Noriega-Papaqui}, {Pelayo},
  {Pretz}, {P{\'e}rez-P{\'e}rez}, {Ren}, {Rho}, {Rivi{\`e}re},
  {Rosa-Gonz{\'a}lez}, {Rosenberg}, {Ruiz-Velasco}, {Salazar}, {Salesa Greus},
  {Sandoval}, {Schneider}, {Schoorlemmer}, {Sinnis}, {Smith}, {Springer},
  {Surajbali}, {Taboada}, {Tibolla}, {Tollefson}, {Torres}, {Ukwatta},
  {Vianello}, {Villase{\~n}or}, {Weisgarber}, {Westerhoff}, {Wisher}, {Wood},
  {Yapici}, {Younk}, {Zepeda}, \& {Zhou}}]{2017ApJ...843...40A}
{Abeysekara}, A.~U., {Albert}, A., {Alfaro}, R., {et~al.} 2017, \apj, 843, 40

\bibitem[{{Abeysekara} {et~al.}(2021){Abeysekara}, {Albert}, {Alfaro},
  {Alvarez}, {Camacho}, {Arteaga-Vel{\'a}zquez}, {Arunbabu}, {Rojas},
  {Solares}, {Baghmanyan}, {Belmont-Moreno}, {BenZvi}, {Blandford}, {Brisbois},
  {Caballero-Mora}, {Capistr{\'a}n}, {Carrami{\~n}ana}, {Casanova}, {Cotti},
  {Le{\'o}n}, {De la Fuente}, {Hernandez}, {Dingus}, {DuVernois}, {Durocher},
  {D{\'\i}az-V{\'e}lez}, {Ellsworth}, {Engel}, {Espinoza}, {Fan}, {Fang},
  {Fleischhack}, {Fraija}, {Galv{\'a}n-G{\'a}mez}, {Garcia},
  {Garc{\'\i}a-Gonz{\'a}lez}, {Garfias}, {Giacinti}, {Gonz{\'a}lez}, {Goodman},
  {Harding}, {Hernandez}, {Hinton}, {Hona}, {Huang}, {Hueyotl-Zahuantitla},
  {H{\"u}ntemeyer}, {Iriarte}, {Jardin-Blicq}, {Joshi}, {Kieda}, {Lara}, {Lee},
  {Vargas}, {Linnemann}, {Longinotti}, {Luis-Raya}, {Lundeen}, {Malone},
  {Martinez}, {Martinez-Castellanos}, {Mart{\'\i}nez-Castro}, {Matthews},
  {Miranda-Romagnoli}, {Morales-Soto}, {Moreno}, {Mostaf{\'a}}, {Nayerhoda},
  {Nellen}, {Newbold}, {Nisa}, {Noriega-Papaqui}, {Olivera-Nieto}, {Omodei},
  {Peisker}, {P{\'e}rez Araujo}, {P{\'e}rez-P{\'e}rez}, {Ren}, {Rho},
  {Rosa-Gonz{\'a}lez}, {Ruiz-Velasco}, {Salazar}, {Greus}, {Sandoval},
  {Schneider}, {Schoorlemmer}, {Serna}, {Smith}, {Springer}, {Surajbali},
  {Tollefson}, {Torres}, {Torres-Escobedo}, {Ure{\~n}a-Mena}, {Weisgarber},
  {Werner}, {Willox}, {Zepeda}, {Zhou}, {De Le{\'o}n}, \&
  {{\'A}lvarez}}]{2021NatAs...5..465A}
{Abeysekara}, A.~U., {Albert}, A., {Alfaro}, R., {et~al.} 2021, Nature
  Astronomy, 5, 465

\bibitem[{{Aharonian} {et~al.}(2008){Aharonian}, {Akhperjanian}, {Bazer-Bachi},
  {Behera}, {Beilicke}, {Benbow}, {Berge}, {Bernl{\"o}hr}, {Boisson}, {Bolz},
  {Borrel}, {Braun}, {Brion}, {Brown}, {B{\"u}hler}, {Bulik}, {B{\"u}sching},
  {Boutelier}, {Carrigan}, {Chadwick}, {Chounet}, {Clapson}, {Coignet},
  {Cornils}, {Costamante}, {Degrange}, {Dickinson}, {Djannati-Ata{\"\i}},
  {Domainko}, {O'C. Drury}, {Dubus}, {Dyks}, {Egberts}, {Emmanoulopoulos},
  {Espigat}, {Farnier}, {Feinstein}, {Fiasson}, {F{\"o}rster}, {Fontaine},
  {Fukui}, {Funk}, {Funk}, {F{\"u}{\ss}ling}, {Gallant}, {Giebels},
  {Glicenstein}, {Gl{\"u}ck}, {Goret}, {Hadjichristidis}, {Hauser}, {Hauser},
  {Heinzelmann}, {Henri}, {Hermann}, {Hinton}, {Hoffmann}, {Hofmann},
  {Holleran}, {Hoppe}, {Horns}, {Jacholkowska}, {de Jager}, {Kendziorra},
  {Kerschhaggl}, {Kh{\'e}lifi}, {Komin}, {Kosack}, {Lamanna}, {Latham}, {Le
  Gallou}, {Lemi{\`e}re}, {Lemoine-Goumard}, {Lenain}, {Lohse}, {Martin},
  {Martineau-Huynh}, {Marcowith}, {Masterson}, {Maurin}, {McComb}, {Moderski},
  {Moriguchi}, {Moulin}, {de Naurois}, {Nedbal}, {Nolan}, {Olive}, {Orford},
  {Osborne}, {Ostrowski}, {Panter}, {Pedaletti}, {Pelletier}, {Petrucci},
  {Pita}, {P{\"u}hlhofer}, {Punch}, {Ranchon}, {Raubenheimer}, {Raue},
  {Rayner}, {Reimer}, {Renaud}, {Ripken}, {Rob}, {Rolland}, {Rosier-Lees},
  {Rowell}, {Rudak}, {Ruppel}, {Sahakian}, {Santangelo}, {Saug{\'e}},
  {Schlenker}, {Schlickeiser}, {Schr{\"o}der}, {Schwanke}, {Schwarzburg},
  {Schwemmer}, {Shalchi}, {Sol}, {Spangler}, {Stawarz}, {Steenkamp},
  {Stegmann}, {Superina}, {Takeuchi}, {Tam}, {Tavernet}, {Terrier}, {van
  Eldik}, {Vasileiadis}, {Venter}, {Vialle}, {Vincent}, {Vivier}, {V{\"o}lk},
  {Volpe}, {Wagner}, \& {Ward}}]{aharonian2008discovery}
{Aharonian}, F., {Akhperjanian}, A.~G., {Bazer-Bachi}, A.~R., {et~al.} 2008,
  \aap, 481, 401

\bibitem[{{Aharonian} {et~al.}(2004){Aharonian}, {Akhperjanian}, {Aye},
  {Bazer-Bachi}, {Beilicke}, {Benbow}, {Berge}, {Berghaus}, {Bernl{\"o}hr},
  {Bolz}, {Boisson}, {Borgmeier}, {Breitling}, {Brown}, {Bussons Gordo},
  {Chadwick}, {Chitnis}, {Chounet}, {Cornils}, {Costamante}, {Degrange},
  {Djannati-Ata{\"\i}}, {Drury}, {Ergin}, {Espigat}, {Feinstein}, {Fleury},
  {Fontaine}, {Funk}, {Gallant}, {Giebels}, {Gillessen}, {Goret}, {Guy},
  {Hadjichristidis}, {Hauser}, {Heinzelmann}, {Henri}, {Hermann}, {Hinton},
  {Hofmann}, {Holleran}, {Horns}, {de Jager}, {Jung}, {Kh{\'e}lifi}, {Komin},
  {Konopelko}, {Latham}, {Le Gallou}, {Lemoine}, {Lemi{\`e}re}, {Leroy},
  {Lohse}, {Marcowith}, {Masterson}, {McComb}, {de Naurois}, {Nolan},
  {Noutsos}, {Orford}, {Osborne}, {Ouchrif}, {Panter}, {Pelletier}, {Pita},
  {Pohl}, {P{\"u}hlhofer}, {Punch}, {Raubenheimer}, {Raue}, {Raux}, {Rayner},
  {Redondo}, {Reimer}, {Reimer}, {Ripken}, {Rivoal}, {Rob}, {Rolland},
  {Rowell}, {Sahakian}, {Saug{\'e}}, {Schlenker}, {Schlickeiser}, {Schuster},
  {Schwanke}, {Siewert}, {Sol}, {Steenkamp}, {Stegmann}, {Tavernet},
  {Th{\'e}oret}, {Tluczykont}, {van der Walt}, {Vasileiadis}, {Vincent},
  {Visser}, {V{\"o}lk}, \& {Wagner}}]{aharonian2004}
{Aharonian}, F.~A., {Akhperjanian}, A.~G., {Aye}, K.~M., {et~al.} 2004, \nat,
  432, 75

\bibitem[{{Aharonian} \& {Atoyan}(1996)}]{aharonian1996emissivity}
{Aharonian}, F.~A. \& {Atoyan}, A.~M. 1996, \aap, 309, 917

\bibitem[{{Akaike}(1974)}]{1974AIC}
{Akaike}, H. 1974, IEEE Transactions on Automatic Control, 19, 716

\bibitem[{{Atwood} {et~al.}(2009){Atwood}, {Abdo}, {Ackermann}, {Althouse},
  {Anderson}, {Axelsson}, {Baldini}, {Ballet}, {Band}, {Barbiellini},
  {Bartelt}, {Bastieri}, {Baughman}, {Bechtol}, {B{\'e}d{\'e}r{\`e}de},
  {Bellardi}, {Bellazzini}, {Berenji}, {Bignami}, {Bisello}, {Bissaldi},
  {Blandford}, {Bloom}, {Bogart}, {Bonamente}, {Bonnell}, {Borgland},
  {Bouvier}, {Bregeon}, {Brez}, {Brigida}, {Bruel}, {Burnett}, {Busetto},
  {Caliandro}, {Cameron}, {Caraveo}, {Carius}, {Carlson}, {Casandjian},
  {Cavazzuti}, {Ceccanti}, {Cecchi}, {Charles}, {Chekhtman}, {Cheung},
  {Chiang}, {Chipaux}, {Cillis}, {Ciprini}, {Claus}, {Cohen-Tanugi},
  {Condamoor}, {Conrad}, {Corbet}, {Corucci}, {Costamante}, {Cutini}, {Davis},
  {Decotigny}, {DeKlotz}, {Dermer}, {de Angelis}, {Digel}, {do Couto e Silva},
  {Drell}, {Dubois}, {Dumora}, {Edmonds}, {Fabiani}, {Farnier}, {Favuzzi},
  {Flath}, {Fleury}, {Focke}, {Funk}, {Fusco}, {Gargano}, {Gasparrini},
  {Gehrels}, {Gentit}, {Germani}, {Giebels}, {Giglietto}, {Giommi}, {Giordano},
  {Glanzman}, {Godfrey}, {Grenier}, {Grondin}, {Grove}, {Guillemot}, {Guiriec},
  {Haller}, {Harding}, {Hart}, {Hays}, {Healey}, {Hirayama}, {Hjalmarsdotter},
  {Horn}, {Hughes}, {J{\'o}hannesson}, {Johansson}, {Johnson}, {Johnson},
  {Johnson}, {Johnson}, {Kamae}, {Katagiri}, {Kataoka}, {Kavelaars}, {Kawai},
  {Kelly}, {Kerr}, {Klamra}, {Kn{\"o}dlseder}, {Kocian}, {Komin}, {Kuehn},
  {Kuss}, {Landriu}, {Latronico}, {Lee}, {Lee}, {Lemoine-Goumard}, {Lionetto},
  {Longo}, {Loparco}, {Lott}, {Lovellette}, {Lubrano}, {Madejski}, {Makeev},
  {Marangelli}, {Massai}, {Mazziotta}, {McEnery}, {Menon}, {Meurer},
  {Michelson}, {Minuti}, {Mirizzi}, {Mitthumsiri}, {Mizuno}, {Moiseev},
  {Monte}, {Monzani}, {Moretti}, {Morselli}, {Moskalenko}, {Murgia},
  {Nakamori}, {Nishino}, {Nolan}, {Norris}, {Nuss}, {Ohno}, {Ohsugi}, {Omodei},
  {Orlando}, {Ormes}, {Paccagnella}, {Paneque}, {Panetta}, {Parent}, {Pearce},
  {Pepe}, {Perazzo}, {Pesce-Rollins}, {Picozza}, {Pieri}, {Pinchera}, {Piron},
  {Porter}, {Poupard}, {Rain{\`o}}, {Rando}, {Rapposelli}, {Razzano}, {Reimer},
  {Reimer}, {Reposeur}, {Reyes}, {Ritz}, {Rochester}, {Rodriguez}, {Romani},
  {Roth}, {Russell}, {Ryde}, {Sabatini}, {Sadrozinski}, {Sanchez}, {Sander},
  {Sapozhnikov}, {Parkinson}, {Scargle}, {Schalk}, {Scolieri}, {Sgr{\`o}},
  {Share}, {Shaw}, {Shimokawabe}, {Shrader}, {Sierpowska-Bartosik}, {Siskind},
  {Smith}, {Smith}, {Spandre}, {Spinelli}, {Starck}, {Stephens}, {Strickman},
  {Strong}, {Suson}, {Tajima}, {Takahashi}, {Takahashi}, {Tanaka}, {Tenze},
  {Tether}, {Thayer}, {Thayer}, {Thompson}, {Tibaldo}, {Tibolla}, {Torres},
  {Tosti}, {Tramacere}, {Turri}, {Usher}, {Vilchez}, {Vitale}, {Wang},
  {Watters}, {Winer}, {Wood}, {Ylinen}, \& {Ziegler}}]{Atwood2009}
{Atwood}, W.~B., {Abdo}, A.~A., {Ackermann}, M., {et~al.} 2009, \apj, 697, 1071

\bibitem[{{Barr} {et~al.}(2013){Barr}, {Guillemot}, {Champion}, {Kramer},
  {Eatough}, {Lee}, {Verbiest}, {Bassa}, {Camilo}, {{\c{C}}elik}, {Cognard},
  {Ferrara}, {Freire}, {Janssen}, {Johnston}, {Keith}, {Lyne}, {Michelson},
  {Parkinson}, {Ransom}, {Ray}, {Stappers}, \& {Wood}}]{2013MNRAS.429.1633B}
{Barr}, E.~D., {Guillemot}, L., {Champion}, D.~J., {et~al.} 2013, \mnras, 429,
  1633

\bibitem[{{Bell}(2004)}]{2004MNRAS.353..550B}
{Bell}, A.~R. 2004, \mnras, 353, 550

\bibitem[{{Bolatto} {et~al.}(2013){Bolatto}, {Wolfire}, \&
  {Leroy}}]{bolatto2013}
{Bolatto}, A.~D., {Wolfire}, M., \& {Leroy}, A.~K. 2013, \araa, 51, 207

\bibitem[{{Cao} {et~al.}(2021{\natexlab{a}}){Cao}, {Aharonian}, {An},
  {Axikegu}, {Bai}, {Bai}, {Bao}, {Bastieri}, {Bi}, {Bi}, {Cai}, {Cai}, {Cao},
  {Chang}, {Chang}, {Chen}, {Chen}, {Chen}, {Chen}, {Chen}, {Chen}, {Chen},
  {Chen}, {Chen}, {Chen}, {Chen}, {Chen}, {Chen}, {Cheng}, {Cheng}, {Cui},
  {Cui}, {Cui}, {Piazzoli}, {Dai}, {Dai}, {Dai}, {Dan-Zeng-Luo-Bu}, {Volpe},
  {Dong}, {Duan}, {Fan}, {Fan}, {Fan}, {Fang}, {Fang}, {Feng}, {Feng}, {Feng},
  {Feng}, {Gao}, {Gao}, {Gao}, {Gao}, {Gao}, {Ge}, {Geng}, {Gong}, {Gou}, {Gu},
  {Guo}, {Guo}, {Guo}, {Guo}, {Guo}, {Han}, {He}, {He}, {He}, {He}, {He}, {He},
  {Heller}, {Hor}, {Hou}, {Hu}, {Hu}, {Hu}, {Hu}, {Huang}, {Huang}, {Huang},
  {Huang}, {Huang}, {Huang}, {Ji}, {Ji}, {Jia}, {Jiang}, {Jiang}, {Jin}, {Ke},
  {Kuleshov}, {Levochkin}, {Li}, {Li}, {Li}, {Li}, {Li}, {Li}, {Li}, {Li},
  {Li}, {Li}, {Li}, {Li}, {Li}, {Li}, {Li}, {Li}, {Li}, {Liang}, {Liang},
  {Lin}, {Liu}, {Liu}, {Liu}, {Liu}, {Liu}, {Liu}, {Liu}, {Liu}, {Liu}, {Liu},
  {Liu}, {Liu}, {Liu}, {Liu}, {Liu}, {Liu}, {Long}, {Lu}, {Lv}, {Ma}, {Ma},
  {Ma}, {Mao}, {Masood}, {Min}, {Mitthumsiri}, {Montaruli}, {Nan}, {Pang},
  {Pattarakijwanich}, {Pei}, {Qi}, {Qi}, {Qiao}, {Qin}, {Ruffolo}, {Rulev},
  {S{\'a}iz}, {Shao}, {Shchegolev}, {Sheng}, {Shi}, {Song}, {Stenkin},
  {Stepanov}, {Su}, {Sun}, {Sun}, {Sun}, {Tam}, {Tang}, {Tian}, {Wang}, {Wang},
  {Wang}, {Wang}, {Wang}, {Wang}, {Wang}, {Wang}, {Wang}, {Wang}, {Wang},
  {Wang}, {Wang}, {Wang}, {Wang}, {Wang}, {Wang}, {Wang}, {Wang}, {Wang},
  {Wang}, {Wang}, {Wei}, {Wei}, {Wei}, {Wen}, {Wu}, {Wu}, {Wu}, {Wu}, {Wu},
  {Xi}, {Xia}, {Xia}, {Xiang}, {Xiao}, {Xiao}, {Xiao}, {Xin}, {Xin}, {Xing},
  {Xu}, {Xu}, {Xue}, {Yan}, {Yan}, {Yang}, {Yang}, {Yang}, {Yang}, {Yang},
  {Yang}, {Yang}, {Yao}, {Yao}, {Ye}, {Yin}, {Yin}, {You}, {You}, {Yu}, {Yuan},
  {Zeng}, {Zeng}, {Zeng}, {Zeng}, {Zha}, {Zhai}, {Zhang}, {Zhang}, {Zhang},
  {Zhang}, {Zhang}, {Zhang}, {Zhang}, {Zhang}, {Zhang}, {Zhang}, {Zhang},
  {Zhang}, {Zhang}, {Zhang}, {Zhang}, {Zhang}, {Zhang}, {Zhang}, {Zhang},
  {Zhao}, {Zhao}, {Zhao}, {Zhao}, {Zhao}, {Zheng}, {Zheng}, {Zhou}, {Zhou},
  {Zhou}, {Zhou}, {Zhou}, {Zhou}, {Zhu}, {Zhu}, {Zhu}, {Zhu}, \&
  {Zuo}}]{2021ApJ...919L..22C}
{Cao}, Z., {Aharonian}, F., {An}, Q., {et~al.} 2021{\natexlab{a}}, \apjl, 919,
  L22

\bibitem[{{Cao} {et~al.}(2023){Cao}, {Aharonian}, {An}, {Axikegu}, {Bai},
  {Bao}, {Bastieri}, {Bi}, {Bi}, {Cai}, {Cao}, {Cao}, {Cao}, {Chang}, {Chang},
  {Chen}, {Chen}, {Chen}, {Chen}, {Chen}, {Chen}, {Chen}, {Chen}, {Chen},
  {Chen}, {Chen}, {Chen}, {Cheng}, {Cheng}, {Cui}, {Cui}, {Cui}, {Cui}, {Dai},
  {Dai}, {Dai}, {Danzengluobu}, {della Volpe}, {Dong}, {Duan}, {Fan}, {Fan},
  {Fang}, {Fang}, {Feng}, {Feng}, {Feng}, {Feng}, {Feng}, {Gabici}, {Gao},
  {Gao}, {Gao}, {Gao}, {Gao}, {Gao}, {Ge}, {Geng}, {Giacinti}, {Gong}, {Gou},
  {Gu}, {Guo}, {Guo}, {Guo}, {Guo}, {Han}, {He}, {He}, {He}, {He}, {He},
  {Heller}, {Hor}, {Hou}, {Hou}, {Hou}, {Hu}, {Hu}, {Hu}, {Huang}, {Huang},
  {Huang}, {Huang}, {Huang}, {Huang}, {Huang}, {Ji}, {Jia}, {Jia}, {Jiang},
  {Jiang}, {Jiang}, {Jin}, {Kang}, {Ke}, {Kuleshov}, {Kurinov}, {Li}, {Li},
  {Li}, {Li}, {Li}, {Li}, {Li}, {Li}, {Li}, {Li}, {Li}, {Li}, {Li}, {Li}, {Li},
  {Li}, {Li}, {Li}, {Li}, {Liang}, {Liang}, {Lin}, {Liu}, {Liu}, {Liu}, {Liu},
  {Liu}, {Liu}, {Liu}, {Liu}, {Liu}, {Liu}, {Liu}, {Liu}, {Liu}, {Liu}, {Lu},
  {Luo}, {Lv}, {Ma}, {Ma}, {Ma}, {Mao}, {Min}, {Mitthumsiri}, {Mu}, {Nan},
  {Neronov}, {Ou}, {Pang}, {Pattarakijwanich}, {Pei}, {Qi}, {Qi}, {Qiao},
  {Qin}, {Ruffolo}, {S{\'a}iz}, {Semikoz}, {Shao}, {Shao}, {Shchegolev},
  {Sheng}, {Shu}, {Song}, {Stenkin}, {Stepanov}, {Su}, {Sun}, {Sun}, {Sun},
  {Tam}, {Tang}, {Tang}, {Tian}, {Wang}, {Wang}, {Wang}, {Wang}, {Wang},
  {Wang}, {Wang}, {Wang}, {Wang}, {Wang}, {Wang}, {Wang}, {Wang}, {Wang},
  {Wang}, {Wang}, {Wang}, {Wang}, {Wang}, {Wang}, {Wang}, {Wei}, {Wei}, {Wei},
  {Wen}, {Wu}, {Wu}, {Wu}, {Wu}, {Wu}, {Xi}, {Xia}, {Xia}, {Xiang}, {Xiao},
  {Xiao}, {Xin}, {Xin}, {Xing}, {Xiong}, {Xu}, {Xu}, {Xu}, {Xu}, {Xue}, {Yan},
  {Yan}, {Yan}, {Yang}, {Yang}, {Yang}, {Yang}, {Yang}, {Yang}, {Yang}, {Yang},
  {Yang}, {Yao}, {Yao}, {Ye}, {Yin}, {Yin}, {You}, {You}, {Yu}, {Yuan}, {Yue},
  {Zeng}, {Zeng}, {Zeng}, {Zha}, {Zhang}, {Zhang}, {Zhang}, {Zhang}, {Zhang},
  {Zhang}, {Zhang}, {Zhang}, {Zhang}, {Zhang}, {Zhang}, {Zhang}, {Zhang},
  {Zhang}, {Zhang}, {Zhang}, {Zhang}, {Zhang}, {Zhao}, {Zhao}, {Zhao}, {Zhao},
  {Zhao}, {Zheng}, {Zhou}, {Zhou}, {Zhou}, {Zhou}, {Zhou}, {Zhou}, {Zhou},
  {Zhu}, {Zhu}, {Zhu}, {Zhu}, \& {Zuo.}}]{2023arXiv230517030C}
{Cao}, Z., {Aharonian}, F., {An}, Q., {et~al.} 2023, arXiv e-prints,
  arXiv:2305.17030

\bibitem[{{Cao} {et~al.}(2021{\natexlab{b}}){Cao}, {Aharonian}, {An},
  {Axikegu}, {Bai}, {Bao}, {Bastieri}, {Bi}, {Bi}, {Cai}, {Cai}, {Cao},
  {Chang}, {Chang}, {Chang}, {Chen}, {Chen}, {Chen}, {Chen}, {Chen}, {Chen},
  {Chen}, {Chen}, {Chen}, {Chen}, {Chen}, {Chen}, {Chen}, {Cheng}, {Cheng},
  {Cui}, {Cui}, {Cui}, {Dai}, {Dai}, {Dai}, {Danzengluobu}, {della Volpe},
  {D'Ettorre Piazzoli}, {Dong}, {Fan}, {Fan}, {Fan}, {Fang}, {Fang}, {Feng},
  {Feng}, {Feng}, {Feng}, {Gao}, {Gao}, {Gao}, {Gao}, {Ge}, {Geng}, {Gong},
  {Gou}, {Gu}, {Guo}, {Guo}, {Guo}, {Guo}, {Han}, {He}, {He}, {He}, {He}, {He},
  {He}, {Heller}, {Hor}, {Hou}, {Hou}, {Hu}, {Hu}, {Hu}, {Hu}, {Huang},
  {Huang}, {Huang}, {Huang}, {Huang}, {Ji}, {Ji}, {Jia}, {Jiang}, {Jiang},
  {Jin}, {Kuleshov}, {Levochkin}, {Li}, {Li}, {Li}, {Li}, {Li}, {Li}, {Li},
  {Li}, {Li}, {Li}, {Li}, {Li}, {Li}, {Li}, {Li}, {Li}, {Li}, {Liang}, {Liang},
  {Lin}, {Liu}, {Liu}, {Liu}, {Liu}, {Liu}, {Liu}, {Liu}, {Liu}, {Liu}, {Liu},
  {Liu}, {Liu}, {Liu}, {Liu}, {Liu}, {Long}, {Lu}, {Lv}, {Ma}, {Ma}, {Ma},
  {Mao}, {Masood}, {Mitthumsiri}, {Montaruli}, {Nan}, {Pang},
  {Pattarakijwanich}, {Pei}, {Qi}, {Ruffolo}, {Rulev}, {S{\'a}iz}, {Shao},
  {Shchegolev}, {Sheng}, {Shi}, {Song}, {Stenkin}, {Stepanov}, {Sun}, {Sun},
  {Sun}, {Tam}, {Tang}, {Tian}, {Wang}, {Wang}, {Wang}, {Wang}, {Wang}, {Wang},
  {Wang}, {Wang}, {Wang}, {Wang}, {Wang}, {Wang}, {Wang}, {Wang}, {Wang},
  {Wang}, {Wang}, {Wang}, {Wang}, {Wang}, {Wang}, {Wei}, {Wei}, {Wei}, {Wen},
  {Wu}, {Wu}, {Wu}, {Wu}, {Wu}, {Xi}, {Xia}, {Xia}, {Xiang}, {Xiao}, {Xiao},
  {Xin}, {Xin}, {Xing}, {Xu}, {Xu}, {Xue}, {Yan}, {Yang}, {Yang}, {Yang},
  {Yang}, {Yang}, {Yang}, {Yang}, {Yao}, {Yao}, {Ye}, {Yin}, {Yin}, {You},
  {You}, {Yu}, {Yuan}, {Zeng}, {Zeng}, {Zeng}, {Zeng}, {Zha}, {Zhai}, {Zhang},
  {Zhang}, {Zhang}, {Zhang}, {Zhang}, {Zhang}, {Zhang}, {Zhang}, {Zhang},
  {Zhang}, {Zhang}, {Zhang}, {Zhang}, {Zhang}, {Zhang}, {Zhang}, {Zhang},
  {Zhang}, {Zhang}, {Zhao}, {Zhao}, {Zhao}, {Zhao}, {Zhao}, {Zheng}, {Zheng},
  {Zhou}, {Zhou}, {Zhou}, {Zhou}, {Zhou}, {Zhou}, {Zhu}, {Zhu}, {Zhu}, {Zhu},
  \& {Zuo}}]{2021Natur.594...33C}
{Cao}, Z., {Aharonian}, F.~A., {An}, Q., {et~al.} 2021{\natexlab{b}}, \nat,
  594, 33

\bibitem[{{Castelletti} {et~al.}(2013){Castelletti}, {Supan}, {Dubner},
  {Joshi}, \& {Surnis}}]{2013A&A...557L..15C}
{Castelletti}, G., {Supan}, L., {Dubner}, G., {Joshi}, B.~C., \& {Surnis},
  M.~P. 2013, \aap, 557, L15

\bibitem[{{Celli} {et~al.}(2019){Celli}, {Morlino}, {Gabici}, \&
  {Aharonian}}]{2019MNRAS.487.3199C}
{Celli}, S., {Morlino}, G., {Gabici}, S., \& {Aharonian}, F.~A. 2019, \mnras,
  487, 3199

\bibitem[{{Cristofari} {et~al.}(2013){Cristofari}, {Gabici}, {Casanova},
  {Terrier}, \& {Parizot}}]{2013MNRAS.434.2748C}
{Cristofari}, P., {Gabici}, S., {Casanova}, S., {Terrier}, R., \& {Parizot}, E.
  2013, \mnras, 434, 2748

\bibitem[{{Cristofari} {et~al.}(2021){Cristofari}, {Niro}, \&
  {Gabici}}]{2021MNRAS.508.2204C}
{Cristofari}, P., {Niro}, V., \& {Gabici}, S. 2021, \mnras, 508, 2204

\bibitem[{{Dame} {et~al.}(2001){Dame}, {Hartmann}, \&
  {Thaddeus}}]{2001ApJ...547..792D}
{Dame}, T.~M., {Hartmann}, D., \& {Thaddeus}, P. 2001, \apj, 547, 792

\bibitem[{{de la Fuente} {et~al.}(2023{\natexlab{a}}){de la Fuente},
  {Toledano-Juarez}, {Kawata}, {Trinidad}, {Tafoya}, {Sano}, {Tokuda},
  {Nishimura}, {Onishi}, {Sako}, {Hona}, {Ohnishi}, \&
  {Takita}}]{2023PASJ...75..546D}
{de la Fuente}, E., {Toledano-Juarez}, I., {Kawata}, K., {et~al.}
  2023{\natexlab{a}}, \pasj, 75, 546

\bibitem[{{de la Fuente} {et~al.}(2023{\natexlab{b}}){de la Fuente},
  {Toledano-Ju{\'a}rez}, {Kawata}, {Trinidad}, {Yamagishi}, {Takekawa},
  {Tafoya}, {Ohnishi}, {Nishimura}, {Kato}, {Sako}, {Takita}, {Sano}, \&
  {Yadav}}]{2023A&A...675L...5D}
{de la Fuente}, E., {Toledano-Ju{\'a}rez}, I., {Kawata}, K., {et~al.}
  2023{\natexlab{b}}, \aap, 675, L5

\bibitem[{{Devin} {et~al.}(2020){Devin}, {Lemoine-Goumard}, {Grondin},
  {Castro}, {Ballet}, {Cohen}, \& {Hewitt}}]{2020A&A...643A..28D}
{Devin}, J., {Lemoine-Goumard}, M., {Grondin}, M.~H., {et~al.} 2020, \aap, 643,
  A28

\bibitem[{{Feng} {et~al.}(2024){Feng}, {Chen}, {Su}, {Sun}, {Zhang}, {Zhou}, \&
  {Guo}}]{2024A&A...686A.305F}
{Feng}, J.-C., {Chen}, X., {Su}, Y., {et~al.} 2024, \aap, 686, A305

\bibitem[{{Gabici} {et~al.}(2009){Gabici}, {Aharonian}, \&
  {Casanova}}]{gabici2009}
{Gabici}, S., {Aharonian}, F.~A., \& {Casanova}, S. 2009, \mnras, 396, 1629

\bibitem[{{Gaensler} \& {Slane}(2006)}]{2006ARA&A..44...17G}
{Gaensler}, B.~M. \& {Slane}, P.~O. 2006, \araa, 44, 17

\bibitem[{{Gaia Collaboration}(2020)}]{2020yCat.1350....0G}
{Gaia Collaboration}. 2020, VizieR Online Data Catalog, I/350

\bibitem[{{Gao} \& {Han}(2014)}]{2014A&A...567A..59G}
{Gao}, X.~Y. \& {Han}, J.~L. 2014, \aap, 567, A59

\bibitem[{{Gao} {et~al.}(2011){Gao}, {Han}, {Reich}, {Reich}, {Sun}, \&
  {Xiao}}]{Gao11}
{Gao}, X.~Y., {Han}, J.~L., {Reich}, W., {et~al.} 2011, \aap, 529, A159

\bibitem[{{Gerbrandt} {et~al.}(2014){Gerbrandt}, {Foster}, {Kothes},
  {Geisb{\"u}sch}, \& {Tung}}]{2014A&A...566A..76G}
{Gerbrandt}, S., {Foster}, T.~J., {Kothes}, R., {Geisb{\"u}sch}, J., \& {Tung},
  A. 2014, \aap, 566, A76

\bibitem[{{Giacinti} {et~al.}(2012){Giacinti}, {Kachelrie{\ss}}, \&
  {Semikoz}}]{2012PhRvL.108z1101G}
{Giacinti}, G., {Kachelrie{\ss}}, M., \& {Semikoz}, D.~V. 2012, \prl, 108,
  261101

\bibitem[{{Giacinti} {et~al.}(2020){Giacinti}, {Mitchell}, {L{\'o}pez-Coto},
  {Joshi}, {Parsons}, \& {Hinton}}]{2020A&A...636A.113G}
{Giacinti}, G., {Mitchell}, A.~M.~W., {L{\'o}pez-Coto}, R., {et~al.} 2020,
  \aap, 636, A113

\bibitem[{{Ginzburg} \& {Syrovatskii}(1964)}]{Ginzburg&Syrovatskii1964}
{Ginzburg}, V.~L. \& {Syrovatskii}, S.~I. 1964, {The Origin of Cosmic Rays}

\bibitem[{{Green}(2017)}]{2017yCat.7278....0G}
{Green}, D.~A. 2017, VizieR Online Data Catalog, VII/278

\bibitem[{{Hanabata} {et~al.}(2014){Hanabata}, {Katagiri}, {Hewitt}, {Ballet},
  {Fukazawa}, {Fukui}, {Hayakawa}, {Lemoine-Goumard}, {Pedaletti}, {Strong},
  {Torres}, \& {Yamazaki}}]{hanabata2014detailed}
{Hanabata}, Y., {Katagiri}, H., {Hewitt}, J.~W., {et~al.} 2014, \apj, 786, 145

\bibitem[{{He} {et~al.}(2022){He}, {Cui}, {Yeung}, {Tam}, {Zhang}, \&
  {Chen}}]{2022ApJ...928...89H}
{He}, X., {Cui}, Y., {Yeung}, P. K.~H., {et~al.} 2022, \apj, 928, 89

\bibitem[{{Helene}(1983)}]{helene1983}
{Helene}, O. 1983, Nuclear Instruments and Methods in Physics Research, 212,
  319

\bibitem[{{Hillas}(2005)}]{Hillas2005}
{Hillas}, A.~M. 2005, Journal of Physics G Nuclear Physics, 31, R95

\bibitem[{{Inoue} {et~al.}(2012){Inoue}, {Yamazaki}, {Inutsuka}, \&
  {Fukui}}]{2012ApJ...744...71I}
{Inoue}, T., {Yamazaki}, R., {Inutsuka}, S.-i., \& {Fukui}, Y. 2012, \apj, 744,
  71

\bibitem[{Kafexhiu {et~al.}(2014)Kafexhiu, Aharonian, Taylor, \&
  Vila}]{PhysRevD.90.123014}
Kafexhiu, E., Aharonian, F., Taylor, A.~M., \& Vila, G.~S. 2014, Phys. Rev. D,
  90, 123014

\bibitem[{{Kalberla} {et~al.}(2005){Kalberla}, {Burton}, {Hartmann}, {Arnal},
  {Bajaja}, {Morras}, \& {P{\"o}ppel}}]{2005A&A...440..775K}
{Kalberla}, P.~M.~W., {Burton}, W.~B., {Hartmann}, D., {et~al.} 2005, \aap,
  440, 775

\bibitem[{{Katz} \& {Waxman}(2008)}]{2008JCAP...01..018K}
{Katz}, B. \& {Waxman}, E. 2008, \jcap, 2008, 018

\bibitem[{{Kronberger} {et~al.}(2006){Kronberger}, {Teutsch}, {Alessi},
  {Steine}, {Ferrero}, {Graczewski}, {Juchert}, {Patchick}, {Riddle},
  {Saloranta}, {Schoenball}, \& {Watson}}]{2006A&A...447..921K}
{Kronberger}, M., {Teutsch}, P., {Alessi}, B., {et~al.} 2006, \aap, 447, 921

\bibitem[{{Lande} {et~al.}(2012){Lande}, {Ackermann}, {Allafort}, {Ballet},
  {Bechtol}, {Burnett}, {Cohen-Tanugi}, {Drlica-Wagner}, {Funk}, {Giordano},
  {Grondin}, {Kerr}, \& {Lemoine-Goumard}}]{2012ApJ...756....5L}
{Lande}, J., {Ackermann}, M., {Allafort}, A., {et~al.} 2012, \apj, 756, 5

\bibitem[{{Lhaaso Collaboration}(2024)}]{2024SciBu..69..449L}
{Lhaaso Collaboration}. 2024, Science Bulletin, 69, 449

\bibitem[{{Li} \& {Chen}(2010)}]{li2010gamma}
{Li}, H. \& {Chen}, Y. 2010, \mnras, 409, L35

\bibitem[{{Li} {et~al.}(2023{\natexlab{a}}){Li}, {Liu}, \&
  {He}}]{2023ApJ...953..100L}
{Li}, Y., {Liu}, S., \& {He}, Y. 2023{\natexlab{a}}, \apj, 953, 100

\bibitem[{{Li} {et~al.}(2023{\natexlab{b}}){Li}, {Xin}, {Liu}, \&
  {He}}]{2023ApJ...945...21L}
{Li}, Y., {Xin}, Y., {Liu}, S., \& {He}, Y. 2023{\natexlab{b}}, \apj, 945, 21

\bibitem[{{Liu} {et~al.}(2020){Liu}, {Zeng}, {Xin}, \&
  {Zhu}}]{2020ApJ...897L..34L}
{Liu}, S., {Zeng}, H., {Xin}, Y., \& {Zhu}, H. 2020, \apjl, 897, L34

\bibitem[{{Mares} {et~al.}(2021){Mares}, {Lemoine-Goumard}, {Acero}, {Clark},
  {Devin}, {Gabici}, {Gelfand}, {Green}, \& {Grondin}}]{2021ApJ...912..158M}
{Mares}, A., {Lemoine-Goumard}, M., {Acero}, F., {et~al.} 2021, \apj, 912, 158

\bibitem[{{Mattox} {et~al.}(1996){Mattox}, {Bertsch}, {Chiang}, {Dingus},
  {Digel}, {Esposito}, {Fierro}, {Hartman}, {Hunter}, {Kanbach}, {Kniffen},
  {Lin}, {Macomb}, {Mayer-Hasselwander}, {Michelson}, {von Montigny},
  {Mukherjee}, {Nolan}, {Ramanamurthy}, {Schneid}, {Sreekumar}, {Thompson}, \&
  {Willis}}]{mattox1996likelihood}
{Mattox}, J.~R., {Bertsch}, D.~L., {Chiang}, J., {et~al.} 1996, \apj, 461, 396

\bibitem[{{Nolan} {et~al.}(2012){Nolan}, {Abdo}, {Ackermann}, {Ajello},
  {Allafort}, {Antolini}, {Atwood}, {Axelsson}, {Baldini}, {Ballet},
  {Barbiellini}, {Bastieri}, {Bechtol}, {Belfiore}, {Bellazzini}, {Berenji},
  {Bignami}, {Blandford}, {Bloom}, {Bonamente}, {Bonnell}, {Borgland},
  {Bottacini}, {Bouvier}, {Brandt}, {Bregeon}, {Brigida}, {Bruel}, {Buehler},
  {Burnett}, {Buson}, {Caliandro}, {Cameron}, {Campana}, {Ca{\~n}adas},
  {Cannon}, {Caraveo}, {Casandjian}, {Cavazzuti}, {Ceccanti}, {Cecchi},
  {{\c{C}}elik}, {Charles}, {Chekhtman}, {Cheung}, {Chiang}, {Chipaux},
  {Ciprini}, {Claus}, {Cohen-Tanugi}, {Cominsky}, {Conrad}, {Corbet}, {Cutini},
  {D'Ammando}, {Davis}, {de Angelis}, {DeCesar}, {DeKlotz}, {De Luca}, {den
  Hartog}, {de Palma}, {Dermer}, {Digel}, {Silva}, {Drell}, {Drlica-Wagner},
  {Dubois}, {Dumora}, {Enoto}, {Escande}, {Fabiani}, {Falletti}, {Favuzzi},
  {Fegan}, {Ferrara}, {Focke}, {Fortin}, {Frailis}, {Fukazawa}, {Funk},
  {Fusco}, {Gargano}, {Gasparrini}, {Gehrels}, {Germani}, {Giebels},
  {Giglietto}, {Giommi}, {Giordano}, {Giroletti}, {Glanzman}, {Godfrey},
  {Grenier}, {Grondin}, {Grove}, {Guillemot}, {Guiriec}, {Gustafsson},
  {Hadasch}, {Hanabata}, {Harding}, {Hayashida}, {Hays}, {Hill}, {Horan},
  {Hou}, {Hughes}, {Iafrate}, {Itoh}, {J{\'o}hannesson}, {Johnson}, {Johnson},
  {Johnson}, {Johnson}, {Kamae}, {Katagiri}, {Kataoka}, {Katsuta}, {Kawai},
  {Kerr}, {Kn{\"o}dlseder}, {Kocevski}, {Kuss}, {Lande}, {Landriu},
  {Latronico}, {Lemoine-Goumard}, {Lionetto}, {Llena Garde}, {Longo},
  {Loparco}, {Lott}, {Lovellette}, {Lubrano}, {Madejski}, {Marelli}, {Massaro},
  {Mazziotta}, {McConville}, {McEnery}, {Mehault}, {Michelson}, {Minuti},
  {Mitthumsiri}, {Mizuno}, {Moiseev}, {Mongelli}, {Monte}, {Monzani},
  {Morselli}, {Moskalenko}, {Murgia}, {Nakamori}, {Naumann-Godo}, {Norris},
  {Nuss}, {Nymark}, {Ohno}, {Ohsugi}, {Okumura}, {Omodei}, {Orlando}, {Ormes},
  {Ozaki}, {Paneque}, {Panetta}, {Parent}, {Perkins}, {Pesce-Rollins},
  {Pierbattista}, {Pinchera}, {Piron}, {Pivato}, {Porter}, {Racusin},
  {Rain{\`o}}, {Rando}, {Razzano}, {Razzaque}, {Reimer}, {Reimer}, {Reposeur},
  {Ritz}, {Rochester}, {Romani}, {Roth}, {Rousseau}, {Ryde}, {Sadrozinski},
  {Salvetti}, {Sanchez}, {Saz Parkinson}, {Sbarra}, {Scargle}, {Schalk},
  {Sgr{\`o}}, {Shaw}, {Shrader}, {Siskind}, {Smith}, {Spandre}, {Spinelli},
  {Stephens}, {Strickman}, {Suson}, {Tajima}, {Takahashi}, {Takahashi},
  {Tanaka}, {Thayer}, {Thayer}, {Thompson}, {Tibaldo}, {Tibolla}, {Tinebra},
  {Tinivella}, {Torres}, {Tosti}, {Troja}, {Uchiyama}, {Vandenbroucke}, {Van
  Etten}, {Van Klaveren}, {Vasileiou}, {Vianello}, {Vitale}, {Waite},
  {Wallace}, {Wang}, {Werner}, {Winer}, {Wood}, {Wood}, {Wood}, {Yang}, \&
  {Zimmer}}]{2012ApJS..199...31N}
{Nolan}, P.~L., {Abdo}, A.~A., {Ackermann}, M., {et~al.} 2012, \apjs, 199, 31

\bibitem[{{Ohira} {et~al.}(2011){Ohira}, {Murase}, \& {Yamazaki}}]{ohira2011}
{Ohira}, Y., {Murase}, K., \& {Yamazaki}, R. 2011, \mnras, 410, 1577

\bibitem[{{Paturel} {et~al.}(2003){Paturel}, {Petit}, {Prugniel}, {Theureau},
  {Rousseau}, {Brouty}, {Dubois}, \& {Cambr{\'e}sy}}]{2003A&A...412...45P}
{Paturel}, G., {Petit}, C., {Prugniel}, P., {et~al.} 2003, \aap, 412, 45

\bibitem[{{Peron} {et~al.}(2020){Peron}, {Aharonian}, {Casanova}, {Zanin}, \&
  {Romoli}}]{peron2020gamma}
{Peron}, G., {Aharonian}, F., {Casanova}, S., {Zanin}, R., \& {Romoli}, C.
  2020, \apjl, 896, L23

\bibitem[{{Porter} {et~al.}(2006){Porter}, {Moskalenko}, \&
  {Strong}}]{2006ApJ...648L..29P}
{Porter}, T.~A., {Moskalenko}, I.~V., \& {Strong}, A.~W. 2006, \apjl, 648, L29

\bibitem[{{Porter} {et~al.}(2008){Porter}, {Moskalenko}, {Strong}, {Orlando},
  \& {Bouchet}}]{2008ApJ...682..400P}
{Porter}, T.~A., {Moskalenko}, I.~V., {Strong}, A.~W., {Orlando}, E., \&
  {Bouchet}, L. 2008, \apj, 682, 400

\bibitem[{{Principe} {et~al.}(2020){Principe}, {Mitchell}, {Caroff}, {Hinton},
  {Parsons}, \& {Funk}}]{2020A&A...640A..76P}
{Principe}, G., {Mitchell}, A.~M.~W., {Caroff}, S., {et~al.} 2020, \aap, 640,
  A76

\bibitem[{{Reich} {et~al.}(1997){Reich}, {Reich}, \& {Furst}}]{Reich97}
{Reich}, P., {Reich}, W., \& {Furst}, E. 1997, \aaps, 126, 413

\bibitem[{{Reville} \& {Bell}(2013)}]{2013MNRAS.430.2873R}
{Reville}, B. \& {Bell}, A.~R. 2013, \mnras, 430, 2873

\bibitem[{{Rodriguez Marrero} {et~al.}(2008){Rodriguez Marrero}, {Torres}, {de
  Cea del Pozo}, {Reimer}, \& {Cillis}}]{marrero2008}
{Rodriguez Marrero}, A.~Y., {Torres}, D.~F., {de Cea del Pozo}, E., {Reimer},
  O., \& {Cillis}, A.~N. 2008, \apj, 689, 213

\bibitem[{{Sano} {et~al.}(2013){Sano}, {Tanaka}, {Torii}, {Fukuda}, {Yoshiike},
  {Sato}, {Horachi}, {Kuwahara}, {Hayakawa}, {Matsumoto}, {Inoue}, {Yamazaki},
  {Inutsuka}, {Kawamura}, {Tachihara}, {Yamamoto}, {Okuda}, {Mizuno}, {Onishi},
  {Mizuno}, \& {Fukui}}]{2013ApJ...778...59S}
{Sano}, H., {Tanaka}, T., {Torii}, K., {et~al.} 2013, \apj, 778, 59

\bibitem[{{Sofue} \& {Reich}(1979)}]{Sofue79}
{Sofue}, Y. \& {Reich}, W. 1979, \aaps, 38, 251

\bibitem[{{Su} {et~al.}(2017){Su}, {Zhou}, {Yang}, {Chen}, {Chen}, {Gong}, \&
  {Zhang}}]{2017ApJ...845...48S}
{Su}, Y., {Zhou}, X., {Yang}, J., {et~al.} 2017, \apj, 845, 48

\bibitem[{{Tanaka} {et~al.}(2020){Tanaka}, {Uchida}, {Sano}, \&
  {Tsuru}}]{2020ApJ...900L...5T}
{Tanaka}, T., {Uchida}, H., {Sano}, H., \& {Tsuru}, T.~G. 2020, \apjl, 900, L5

\bibitem[{{Thoudam} \& {H{\"o}randel}(2012)}]{2012MNRAS.419..624T}
{Thoudam}, S. \& {H{\"o}randel}, J.~R. 2012, \mnras, 419, 624

\bibitem[{{Tibet AS{\ensuremath{\gamma}} Collaboration} {et~al.}(2021){Tibet
  AS{\ensuremath{\gamma}} Collaboration}, {Amenomori}, {Bao}, {Bi}, {Chen},
  {Chen}, {Chen}, {Chen}, {Chen}, {Cirennima}, {Danzengluobu}, {Fang}, {Fang},
  {Feng}, {Feng}, {Feng}, {Gao}, {Gou}, {Guo}, {Guo}, {He}, {He}, {Hibino},
  {Hotta}, {Hu}, {Hu}, {Huang}, {Jia}, {Jiang}, {Jin}, {Kasahara}, {Katayose},
  {Kato}, {Kato}, {Kawata}, {Kihara}, {Ko}, {Kozai}, {Labaciren}, {Li}, {Li},
  {Li}, {Lin}, {Liu}, {Liu}, {Liu}, {Liu}, {Liu}, {Lou}, {Lu}, {Meng},
  {Munakata}, {Nakada}, {Nakamura}, {Nanjo}, {Nishizawa}, {Ohnishi}, {Ohura},
  {Ozawa}, {Qian}, {Qu}, {Saito}, {Sakata}, {Sako}, {Shao}, {Shibata},
  {Shiomi}, {Sugimoto}, {Takano}, {Takita}, {Tan}, {Tateyama}, {Torii},
  {Tsuchiya}, {Udo}, {Wang}, {Wu}, {Xue}, {Yamamoto}, {Yang}, {Yokoe}, {Yuan},
  {Zhai}, {Zhang}, {Zhang}, {Zhang}, {Zhang}, {Zhang}, {Zhang}, {Zhang},
  {Zhao}, \& {Zhaxisangzhu}}]{2021NatAs...5..460T}
{Tibet AS{\ensuremath{\gamma}} Collaboration}, {Amenomori}, M., {Bao}, Y.~W.,
  {et~al.} 2021, Nature Astronomy, 5, 460

\bibitem[{{Uchiyama} {et~al.}(2012){Uchiyama}, {Funk}, {Katagiri}, {Katsuta},
  {Lemoine-Goumard}, {Tajima}, {Tanaka}, \& {Torres}}]{uchiyama2012fermi}
{Uchiyama}, Y., {Funk}, S., {Katagiri}, H., {et~al.} 2012, \apjl, 749, L35

\bibitem[{{Wood} {et~al.}(2017){Wood}, {Caputo}, {Charles}, {Di Mauro},
  {Magill}, {Perkins}, \& {Fermi-LAT Collaboration}}]{2017ICRC...35..824W}
{Wood}, M., {Caputo}, R., {Charles}, E., {et~al.} 2017, in International Cosmic
  Ray Conference, Vol. 301, 35th International Cosmic Ray Conference
  (ICRC2017), 824

\bibitem[{{Xi} {et~al.}(2020){Xi}, {Liu}, {Wang}, {Yang}, {Yuan}, \&
  {Zhang}}]{2020ApJ...896L..33X}
{Xi}, S.-Q., {Liu}, R.-Y., {Wang}, X.-Y., {et~al.} 2020, \apjl, 896, L33

\bibitem[{{Xin} {et~al.}(2017){Xin}, {Guo}, {Liao}, {Yuan}, {Liu}, \&
  {Wei}}]{2017ApJ...843...90X}
{Xin}, Y.-L., {Guo}, X.-L., {Liao}, N.-H., {et~al.} 2017, \apj, 843, 90

\bibitem[{{Yuan} {et~al.}(2018){Yuan}, {Liao}, {Xin}, {Li}, {Fan}, {Zhang},
  {Hu}, \& {Bi}}]{2018ApJ...854L..18Y}
{Yuan}, Q., {Liao}, N.-H., {Xin}, Y.-L., {et~al.} 2018, \apjl, 854, L18

\bibitem[{{Zabalza}(2015)}]{zabalza2015naima}
{Zabalza}, V. 2015, in International Cosmic Ray Conference, Vol.~34, 34th
  International Cosmic Ray Conference (ICRC2015), 922

\bibitem[{Zeng {et~al.}(2023)Zeng, Guo, Wu, Su, Liu, \& Zhang}]{Zeng:2023uvu}
Zeng, H., Guo, Y., Wu, H., {et~al.} 2023, PoS, ICRC2023, 606

\bibitem[{{Zeng} {et~al.}(2019){Zeng}, {Xin}, \& {Liu}}]{2019ApJ...874...50Z}
{Zeng}, H., {Xin}, Y., \& {Liu}, S. 2019, \apj, 874, 50

\bibitem[{{Zeng} {et~al.}(2017){Zeng}, {Xin}, {Liu}, {Jokipii}, {Zhang}, \&
  {Zhang}}]{2017ApJ...834..153Z}
{Zeng}, H., {Xin}, Y., {Liu}, S., {et~al.} 2017, \apj, 834, 153

\bibitem[{{Zeng} {et~al.}(2021){Zeng}, {Xin}, {Zhang}, \&
  {Liu}}]{2021ApJ...910...78Z}
{Zeng}, H., {Xin}, Y., {Zhang}, S., \& {Liu}, S. 2021, \apj, 910, 78

\bibitem[{{Zhang} \& {Chen}(2016)}]{2016ApJ...821...43Z}
{Zhang}, X. \& {Chen}, Y. 2016, \apj, 821, 43

\bibitem[{{Zirakashvili} \& {Aharonian}(2010)}]{2010ApJ...708..965Z}
{Zirakashvili}, V.~N. \& {Aharonian}, F.~A. 2010, \apj, 708, 965

\end{thebibliography}

\newpage
\appendix
\onecolumn
\section*{Appendix A: Possibility of others potential theoretical model for multi-wavelength spectra}
\renewcommand{\thesection}{A\arabic{section}}
In contrast to the preceding discussion, we calculated the hadronic contribution in the northern low gas density region ($\rm{n_{NorthLobe}}$ = $\rm 9 d_{0.8}^{-1} \, cm^{-3}$), also, the contribution of lepton components in the southern region. In this scenario, both the northern and southern regions are dominated by the hybrid scenario. The noteworthy point is that, due to the existence of dense molecular clouds interacting with SNRs in the southern region, the magnetic field strength in this area could be significantly amplified around the clump\citep{2010ApJ...708..965Z,2012ApJ...744...71I} because of both the field compression and the magneto-hydrodynamical (MHD) instabilities that progress in the shock-clump interaction\citep{2019MNRAS.487.3199C}. For the sake of simplicity and clarity in the model, here we adopt the electron spectrum with the format of ECPL in Table \ref{table:SrcN}, doubling the magnetic field strength ($\sim$ 15.24 $\mu$G), then we set the synchrotron cooling timescale equaling to the age of SNR G150.3+4.5 (E$_{\rm e, cut}$ = 1.25 $\times$ 10$^{\rm 7}$ t$_{\rm age; yr}^{-1}$ B$_{\rm \mu G}^{\rm -2}$ TeV), then the cutoff energy in electron spectrum should be reduced by four times($\sim$ 4.1 TeV). The total energy of electrons above 1 GeV is calculated as W$_{\rm e}$ = 2.66 $\times$ 10$^{\rm 46}$ erg, which is nearly one order of magnitude lower than the protons energy in the southern region or the electrons energy in the northern region. This indicates that the leptonic contribution in the SouthLobe can be nearly neglected (Fig. \ref{fig:A} panel left), similar to the hadronic contribution in the NorthLobe (Fig. \ref{fig:A} panel middle). In this scenario, the total contribution added from NorthLobe and SouthLobe could also consistent with $\gamma$-ray observation results (Fig. \ref{fig:A} panel right), suggesting the real magnetic field strength in SouthLobe should be constrained around 2 times than NorthLobe. Since there is an absence of X-ray data points (only the upper limit detected), the synchrotron emission cannot be well constrained. Therefore, more high-resolution X-ray surveys should be conducted in this region.

\renewcommand{\thefigure}{A\arabic{figure}}
\setcounter{figure}{0}
\begin{figure*}[htbp]
    \includegraphics[trim={0 0.cm 0 0}, clip, width=0.3\textwidth]{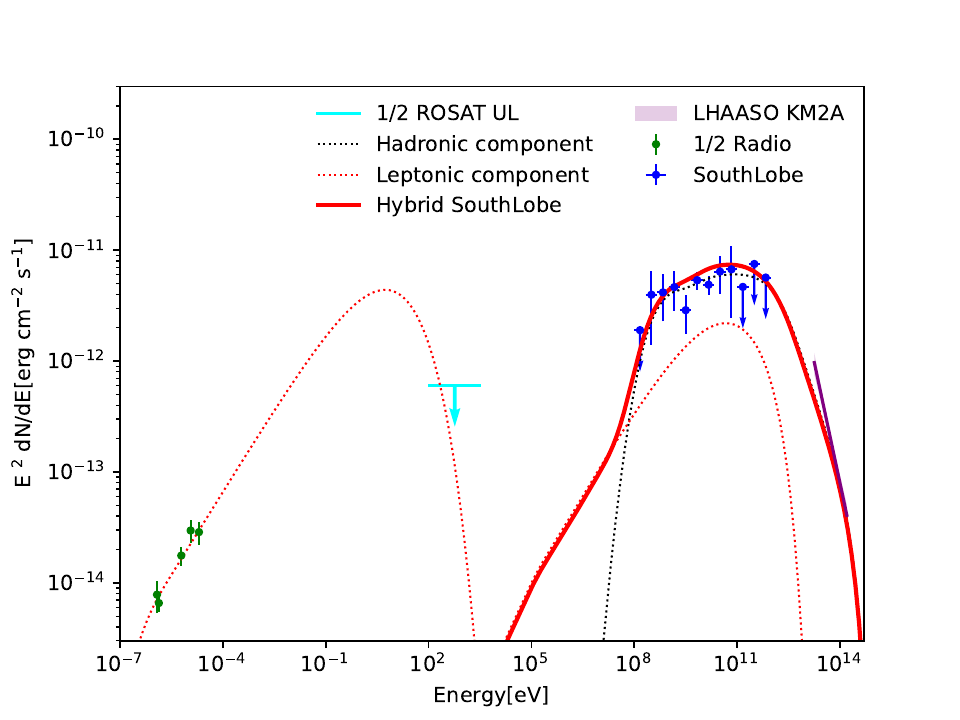}
    \includegraphics[trim={0 0.cm 0 0}, clip, width=0.3\textwidth]{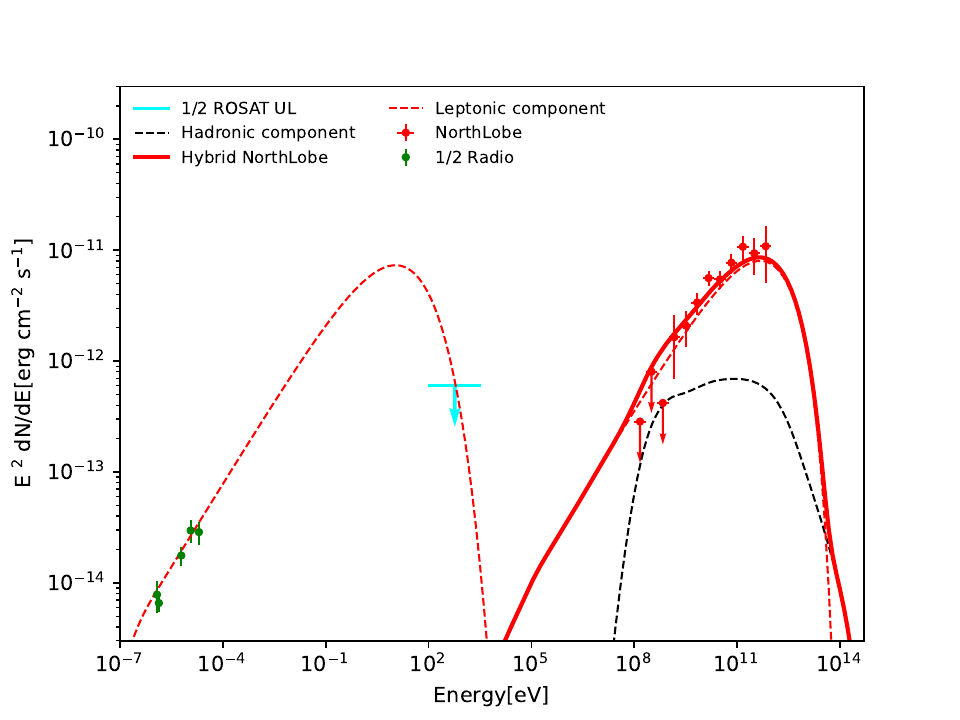}
    \includegraphics[trim={0 0.cm 0 0}, clip, width=0.3\textwidth]{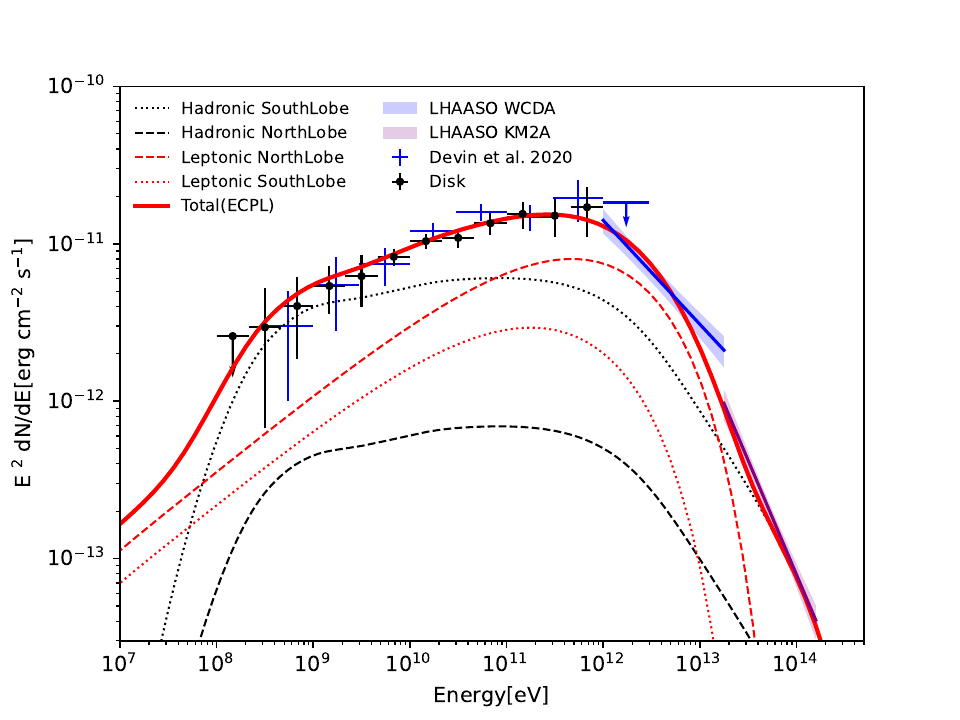}
    \centering
    \caption{Modeling of hybrid scenario in SouthLobe(panel left), NorthLobe (panel middle) and Disk (panel right). In particular, the leptonic contribution in NorthLobe is in agreement with the red dashed line shown in top right panel of Fig. \ref{fig:5}.}
\label{fig:A}
\end{figure*}

\newpage
\renewcommand{\thesection}{B\arabic{section}}
\section*{Appendix B: Chance coincidence with a background $\gamma$-ray source}
Although the above arguments support SrcX origins from the escaped CRs illuminated surrounding molecular clouds, we still cannot totally rule out the possibility that it is a background point-like source just located in the LOS of SNR G150 further spatially coincidence with the surrounding molecular cloud clump. Here we using a Poisson distribution, the chance probability of observing a background point-like source located inside SNR G150.3+4.5 and spatially coincidence with the nearest molecular cloud clump, which can be calculated as $P_{ch}=1-\exp[-\pi (R_0^2+4\sigma_{\gamma}^2+R_c^2) \sum(>F_{th})]$ \citep{2018ApJ...854L..18Y,2020ApJ...896L..33X}, where $\sum(>F_{th})$ is the surface density of sources with fluxes higher than $F_{th}$, $\sigma_{\gamma}$ is the $68\%$ position uncertainties of SrcX, and $R_0$ is the angular distance between the location of SrcX and the target SNR, similarly, $R_c$ is the angular distance between the location of SrcX and surrounding MCs clump. Since the density distribution of the 4FGL sources in the box defined by  Galactic latitude $| b |$ $\textless$ 5$^{\circ}\!$ is uniform with respect to the angle, we estimate a number density of $\sum(>F_{th})=129.5\rm\ sr^{-1}$ (corresponding to 0.04 $\rm\ degree^{-2}$) above the flux  $F_{th}= 1.47 \times 10^{-11} \rm\ erg\,cm^{-2}\,s^{-1}$ (0.1 - 100 GeV), using the power-law  fitting of the cumulative numbers of sources as a function of the threshold fluxes for the 4FGL sources, the chance coincidence probability is estimated to be 0.0018 for $R_0=0^\circ$ (SrcX located inside target SNR and its $\gamma$-ray emission region), $\sigma_{\gamma}=0.025^\circ$ and $R_c  = 0.11^\circ$ (brightest pixel projection distance between the location of SrcX and molecular cloud clump). This results indicate that SrcX, as a background point source, located precisely within the interior of SNR G150 and spatially corresponds to the position of a molecular cloud clump with very low probability.

\renewcommand{\thefigure}{B\arabic{figure}}
\setcounter{figure}{0}
\begin{figure}[htbp]
	\centering
	\begin{minipage}[b]{0.45\textwidth}
		\includegraphics[trim={0 0.cm 0 0}, clip, width=\textwidth]{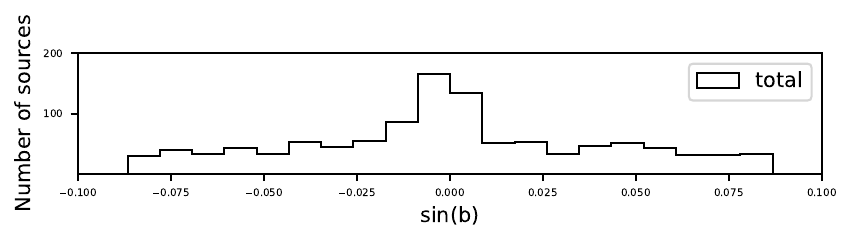}
		\vspace{0.5cm} 
		\includegraphics[trim={0 0.cm 0 0}, clip, width=\textwidth]{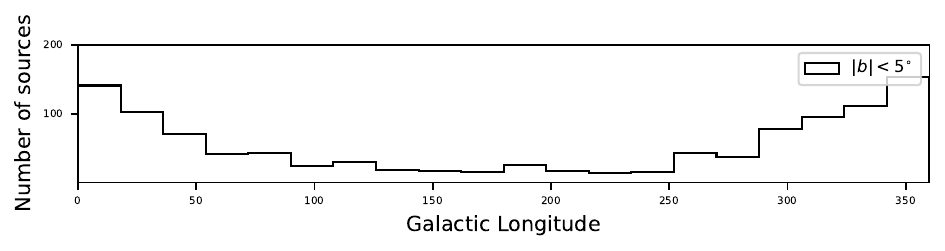}
	\end{minipage}
	\hspace{0.05\textwidth} 
	\begin{minipage}[b]{0.45\textwidth}
		\includegraphics[trim={0 0.cm 0 0}, clip, width=\textwidth]{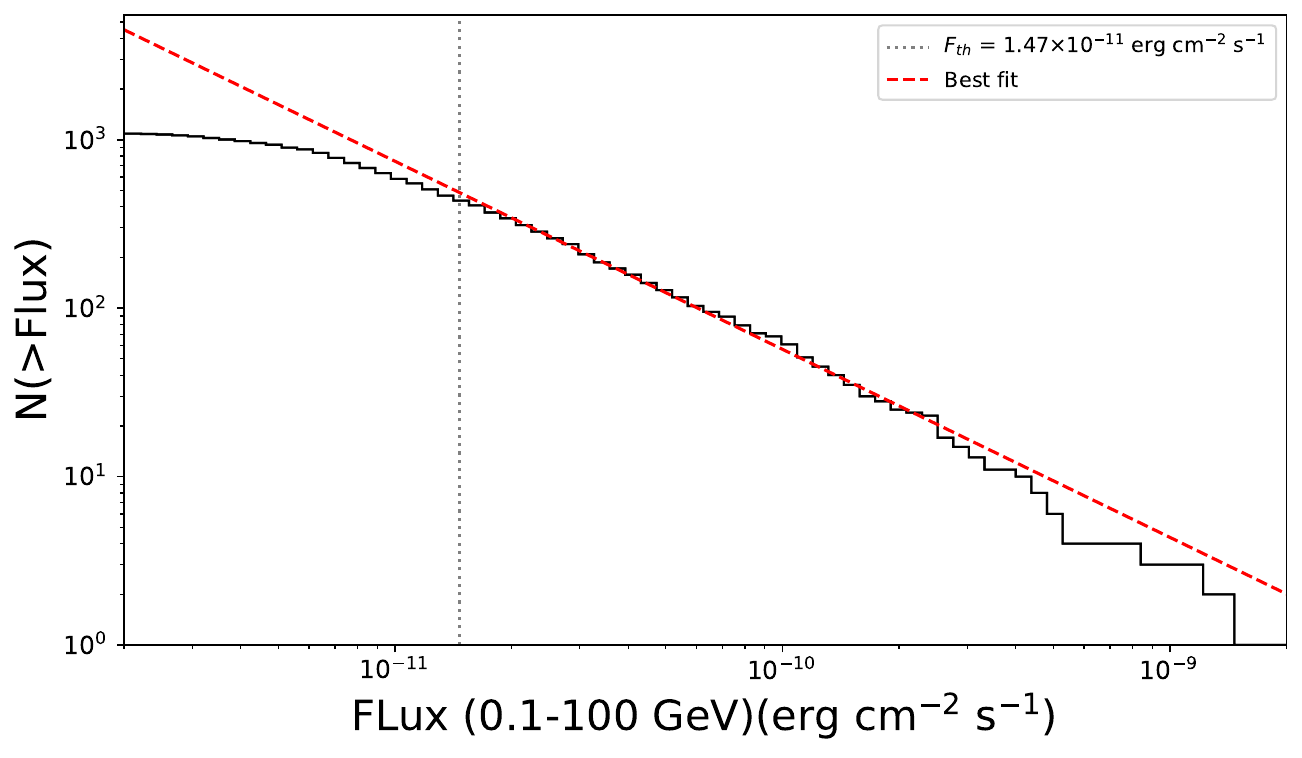}
	\end{minipage}
	\caption{Top left panel: 4FGL source number distribution in galactic latitude ($b$). Bottom left panel: 4FGL source number distribution in galactic longitude for $| b |$ $\textless$ 5$^{\circ}\!$. Right panel: Cumulative numbers of sources and the threshold fluxes for LAT 4FGL sources (histograms) with the power-law fitting results of their high-flux trends (red dashed line). The grey dotted line represents for the 95$\%$ upper limit obtained from flux analysis in Sect. \ref{sec:flux}.}
	\label{fig:B}
\end{figure}

\end{document}